\documentclass[nofootinbib,notitlepage,superscriptaddress,twocolumn]{aastex631}
\pdfoutput=1

\usepackage{natbib}
\usepackage{subfigure}
\usepackage{aas_macros,amsmath,amssymb,esint,graphicx,overpic,xcolor,mathrsfs,stackengine,array,tabularray}
\usepackage[bottom]{footmisc}
\usepackage[]{hyperref,orcidlink}
\usepackage{xspace}

\definecolor{red}{RGB}{228,26,28}
\definecolor{green}{RGB}{77,175,74}
\definecolor{blue}{RGB}{55,126,184}
\definecolor{purple}{RGB}{152,78,163}

\definecolor{Blue}{RGB}{82, 172, 223}
\definecolor{Pink}{RGB}{250, 191, 224}

\definecolor{Blue2}{RGB}{2, 115, 178}
\definecolor{Pink2}{RGB}{204, 122, 168}
\hypersetup{
    colorlinks=true,
    linkcolor=Blue2,        
    urlcolor=Blue2,         
    citecolor=Pink2       
}

\let\originalleft\left
\let\originalright\right
\renewcommand{\left}{\mathopen{}\mathclose\bgroup\originalleft}
\renewcommand{\right}{\aftergroup\egroup\originalright}

\newcommand{\m}{M87*\xspace}

\newcommand{\br}[1]{\left[#1\right]}
\newcommand{\cu}[1]{\left\{#1\right\}}
\newcommand{\pa}[1]{\left(#1\right)}
\newcommand{\ed}{\mathop{}\!\mathrm{d}}

\usepackage{comment}

\begin{document}

\title{Photon Ring Astrometry I:\\ A Simple Spin Measurement Technique for High-Resolution Images of \m}

\author{Delilah E.~A. Gates\,\orcidlink{0000-0002-4882-2674}}
\email{delilah.gates@cfa.harvard.edu}
\affiliation{Center for Astrophysics $\arrowvert$ Harvard \& Smithsonian, 60 Garden Street, Cambridge, MA 02138, USA}
\affiliation{Black Hole Initiative at Harvard University, 20 Garden Street, Cambridge, MA 02138, USA}

\author{Dominic O. Chang\,\orcidlink{0000-0001-9939-5257}}
\affiliation{Center for Astrophysics $\arrowvert$ Harvard \& Smithsonian, 60 Garden Street, Cambridge, MA 02138, USA}
\affiliation{Black Hole Initiative at Harvard University, 20 Garden Street, Cambridge, MA 02138, USA}

\author{Aaron Held\,\orcidlink{0000-0003-2701-9361}}
\affiliation{
Institut de Physique Théorique Philippe Meyer, Laboratoire de Physique de l’\'Ecole normale sup\'erieure (ENS), Universit\'e PSL, CNRS, Sorbonne Universit\'e, Universit\'e Paris Cité, F-75005 Paris, France
}

\author{Daniel~C.~M.~Palumbo\,\orcidlink{0000-0002-7179-3816}}
\affiliation{Black Hole Initiative at Harvard University, 20 Garden Street, Cambridge, MA 02138, USA}
\affiliation{Center for Astrophysics $\arrowvert$ Harvard \& Smithsonian, 60 Garden Street, Cambridge, MA 02138, USA}

\begin{abstract}
The supermassive black hole M87* is a target for future horizon-scale imaging that will resolve the first lensed (${n}~{=}~{1}$) sub-image of the photon ring.
In this work, we identify a concrete observable---the displacement between the centers of the ${n}~{=}~{1}$ photon-ring sub-image and the direct image{.} 
Assuming that the observed large-scale jet of M87 is aligned with the black-hole spin axis, we separate the relative position of the photon ring into components parallel and transverse to the projected spin axis{.} We show that the parallel shift is primarily determined by inclination and emission radius, while the transverse shift is tightly correlated with inclination and spin. 
We demonstrate these effects both in a simple geometric model (to explain the underlying physics) and in {general-relativistic magnetohydrodynamics} simulations with magnetically arrested disks (to provide realistic instantiations). 
We find that a relative astrometric resolution of ${\lesssim}~{0.1\;\mu\rm{as}}${, achievable by the proposed Black Hole Explorer mission,} is sufficient to constrain the spin to better than 9\% if the accretion flow is prograde, {25\% if the flow is retrograde, or 24\% if the direction of the flow is undetermined.}
More generally, this identifies photon-ring astrometry as a promising method to constrain the underlying spacetime geometry without relying on geometric modeling of the observed emission.
\end{abstract}

\section{Introduction}
{Messier 87, first cataloged in 1783, is an elliptical galaxy known today for its prominent jet and central supermassive black hole, \m.}
As of 2019, images of \m have been produced by {the} Event Horizon Telescope Collaboration (EHTC) through the use of very long baseline interferometry (VLBI), a technique that uses an array of telescopes distributed across the globe to synthesize a virtual dish the size of the Earth's diameter {\citep{EHTC_M87_I,EHTC_M87_IV,m87_2018_I,m87_2021}}.
The images all feature a dark central {region} surrounded by a bright ring.

The EHTC's analysis of the \m images from the combined 2017 and 2018 observations shows that the system is well described by synchrotron emission from accreting magnetized plasma.
Their analyses of the total intensity show that a fluid description of the plasma is sufficient for reproducing observations at millimeter wavelengths, with both {strongly-magnetized} magnetically arrested disk (MAD) and {weakly-magnetized} standard and normal evolution (SANE) accretion states as viable possibilities~{\citep{EHTC_M87_V,M872018_theory}}.
Additional constraints from polarized intensity, however, prefer MAD states {\citep{EHTC_M87_VIII}.}
{For details on accretion states, see, e.g., \cite{Narayan_MADs_2003}, \cite{Igumenshchev_MADs_2003}, \cite{Tchekhovskoy_MADs_2011}, \cite{Narayan_SANEs_2012}, and \cite{Skadowski_SANEs_2014}.}
In MAD models, the millimeter-wavelength emission is known to originate predominantly within the first few black-hole gravitational radii and near the equatorial plane~{\citep{Dexter_emission_origin,EHTC_M87_V,Emami_Emissivities_2023}}. 
This property has been used in the past to construct simple geometric ring models that can accurately reproduce the intensity and polarization features seen in images of \m \citep{Narayan_polarized_ring,Gelles_ring}.

M87* is now a key target for space-based VLBI using the Black Hole Explorer (BHEX), which will provide horizon-scale images with improved dynamic range and resolution \citep{bhex_concept, bhex_instrument}. 
{Resolving the horizon-scale image structures will enable constraints of the black hole spin alongside properties of the surrounding accretion flow. Black hole spin is integral to black hole--galaxy co-evolution because it is linked to black hole growth via mergers and accretion and to AGN feedback via launching and powering jets (see, e.g., \citealt{Thorne1974,Blandford1977,Tchekhovskoy_MADs_2011,Volonteri2013,Ricarte2025,Cho2026}.)}

Theory predicts that high-resolution images {feature multiple superposed images of the accretion flow, boasting} a diffuse ring of gently lensed photons formed from direct emission (already resolved in the EHT images) as well as a nested sequence of progressively thinner rings formed from indirect emission, which we collectively refer to as the photon ring. 
{Typically, each image-plane ring (i.e., accretion-flow image) is labeled by an integer ${n}$ denoting the number of half orbits completed by the photons as they travel from the emission locations to the observer: ${n}~{=}~{0}$ for the direct image and ${n}~{=}~{1,2,\dots}$ for the photon-ring sub-images. These successive rings} form a nested ``wedding cake'' feature in black-hole images that converges to a critical curve as ${n}~{\rightarrow}~{\infty}$.
The photon ring is particularly well suited for VLBI observations since it has been shown by \cite{photon_ring_universal} to imprint a universal signature on interferometric measurements. 
{Higher-order $n$ sub-images carry increased information about the spacetime and are less sensitive to the astrophysics of the accretion flow, with the critical curve being completely astrophysics agnostic, determined only by the black hole's properties. As such, the large-$n$ components have been of theoretical interest in the study of black-hole parameter inference.}

There is a growing body of work studying the sensitivity of horizon-{scale} images to black-hole parameters by characterizing features (e.g., size, shape, location, brightness, and polarization pattern) of the critical curve (${n}~{\to}~{\infty}$), the first (${n}~{=}~{1}$) and second (${n}~{=}~{2}$) photon-ring sub-images, and the direct image (${n}~{=}~{0}$) \citep{
Gralla_photon_ring_shape,Himwich_universal_polarimertic,Chael_inner_shadow,Broderick_photon_ring_size,CardenasAvendano_n1_photon_ring_shape,Palumbo_polarimetry_spin,Keeble_n1_photon_ring_shape,Salehi,Walia, Farah_ring_brightness,Kearns_Inner_Shadow,Wong_spin_and_polarization,Bernshteyn_ring_asymmetry, Guan2026_in_prep}.
While planned future observations will not resolve the extremely lensed (${n}~{\geq}~{2}$) emission{, which} well approximates the critical curve, BHEX is expected to resolve the first (${n}~{=}~{1}$) photon-ring sub-image from the direct (${n}~{=}~{0}$) emission. BHEX will also perform repeated observations of \m to determine its average appearance. In particular, BHEX observations can provide precise measurements of the ${n}~{=}~{1}$ sub-image's size and shape and its location relative to the direct image, as encoded in the interferometric amplitude and phase \citep{photon_ring_universal,CardenasAvendano_measuring_n1,GutierrezLara_interferometric_phase}. 

Regarding the relative location of horizon-scale image features, previous authors have suggested that frame-dragging effects can cause a shift in the critical curve that is measurable {when} referenced to a direct-image feature such as the inner shadow \citep{Chael_inner_shadow} or the large-scale jet \citep{jet_photon_ring_shift}.
Here, we will show that the relationship between ${n}~{=}~{0}$ and ${n}~{=}~{1}$ sub-image centers can serve as an accurate proxy for the shifting behavior of the critical curve, as we will describe heuristically in \autoref{sec:setup}.
In \autoref{sec:ERM}, under the assumption of equatorial emission, we will show that the normalized parallel shift of the ${n}~{=}~{1}$ image, with respect to the projection of the black-hole spin axis, is sensitive to inclination and emission radius, while the normalized transverse shift is sensitive to inclination and spin. 
The behavior of the normalized transverse shift suggests that it can be used to uniquely constrain spin if the inclination of the black hole is known, and thus may be useful for spin measurements of \m whose jet orientation is constrained \citep{Walker_jet}.
Furthermore, in \autoref{sec:GRMHD}, we will show that this shift can be measured directly from the {time-averaged images of MAD systems, remaining robust despite uncertainties arising from realistic astrophysical variation.}
{Finally, in \autoref{sec:Discussion}, we derive relative astrometric resolution required for spin measurements of \m and compare relative astrometry to more sophisticated spin-constraining techniques that use geometric emission models.}

\section{Relative center shift of the direct image and photon ring}
\label{sec:setup}

The geometry around a black hole is described by the Kerr metric, which depends on the black-hole mass $M$ and {angular momentum $J$. 
Through the angular momentum, we define the dimensionless spin parameter ${a_*}~{\equiv}~{Jc/(GM^2)}$, where $G$ is the gravitational constant, and $c$ is the speed of light, bounded by the range $a_*\in[0,1]$ where the extremes represent the non-rotating and maximally rotating black hole.}
When the distance to the black hole (${r_o}~{=}~{D}$) is much larger than the black hole's gravitational radius ${r_g}~{=}~{GM/c^2}$, the on-sky appearance of a source location can be characterized by the black hole's mass-to-distance ratio ${\theta_g}~{=}~{GM/({c^2 D})}$ and the arriving photon's four-momentum. We define the image-plane coordinates,
\begin{align}
    \label{eq:ScreenCoords}
    (x,y)=\theta_g \pa{-p_\phi \csc\theta_o,p_\theta},
\end{align}
where $p_\phi$ and $p_\theta$ are the azimuthal and polar components of the photon's four co-momentum normalized by its energy (see \cite{Cunningham_Bardeen} for details).
In these coordinates, the $y$-axis corresponds to the projection of the black hole's spin axis. 
If we assume that the black-hole spin axis is aligned with the jet, then we may use the jet to orient the coordinate system in realistic horizon-scale images of \m.
This assumption also allows us to set {$\theta_o$, the inclination of the observer with respect to the black hole's spin vector,} to be either ${\approx}~{17^\circ}$ or its complement (${\theta_o}~{\approx}~{163^\circ}$) \citep{Walker_jet}, though the observed brightness asymmetry of the \m accretion disk suggests that its inclination is likely $163^\circ$; see \cite{EHTC_M87_V} and \cite{m87_2021}, where{,} for comparison{,} the coordinate system of \autoref{eq:ScreenCoords} should be rotated clockwise by $72^\circ$ to be aligned with the position angle of the large-scale jet. 

In Boyer-Lindquist coordinates $(t,r,\theta,\phi)$, for a given source point $\cu{r_s,\theta_s,\phi_s}$ and the observation point $\cu{r_o,\theta_o,\phi_o}$, photons can take a countably infinite number of paths from the source to observer, with each path accruing a different amount of angular deflection~\citep{Dexter_Kerr_photon_orbits,Kraniotis_grav_lenses,Gralla_Kerr_geodesics}~\footnote{
A source point and the observer may be linked by multiple geodesics that have the same number of angular turning points because the black-hole spacetime admits caustics. These geodesics are considered to be of the same lensing order $n$ \citep{Caustics_Rauch,Caustics_Bozza,subn_ring}{.}}.
The photons form images comprising the most weakly lensed (${n}~{=}~{0}$) sub-image, which appears as a diffuse ring of intensity, and higher-lensed (${n}~{\geq}~{1}$) sub-images, which together form the photon ring{.}
The discrete, sub-image structure of black-hole images is advantageous because we cannot precisely locate the origin of our coordinate system in horizon-scale images. 
However, with a suitable geometric procedure, we can define the direct sub-image center ($x_0,y_0$) and secondary sub-image center $(x_1,y_1)$ up to an overall coordinate offset. Thus, such a geometric procedure allows us to measure the transverse and parallel shifts, $|{x_0}~{-}~{x_1}|$ and $|{y_0}~{-}~{y_1}|$.

\begin{figure*}
    {{\Large \textcolor{white}{I} \hspace{.9cm} MAD Simulation \hspace{3.8cm} Geometric Equatorial Disk}}   \\
    \centering
    \includegraphics[width=\linewidth]{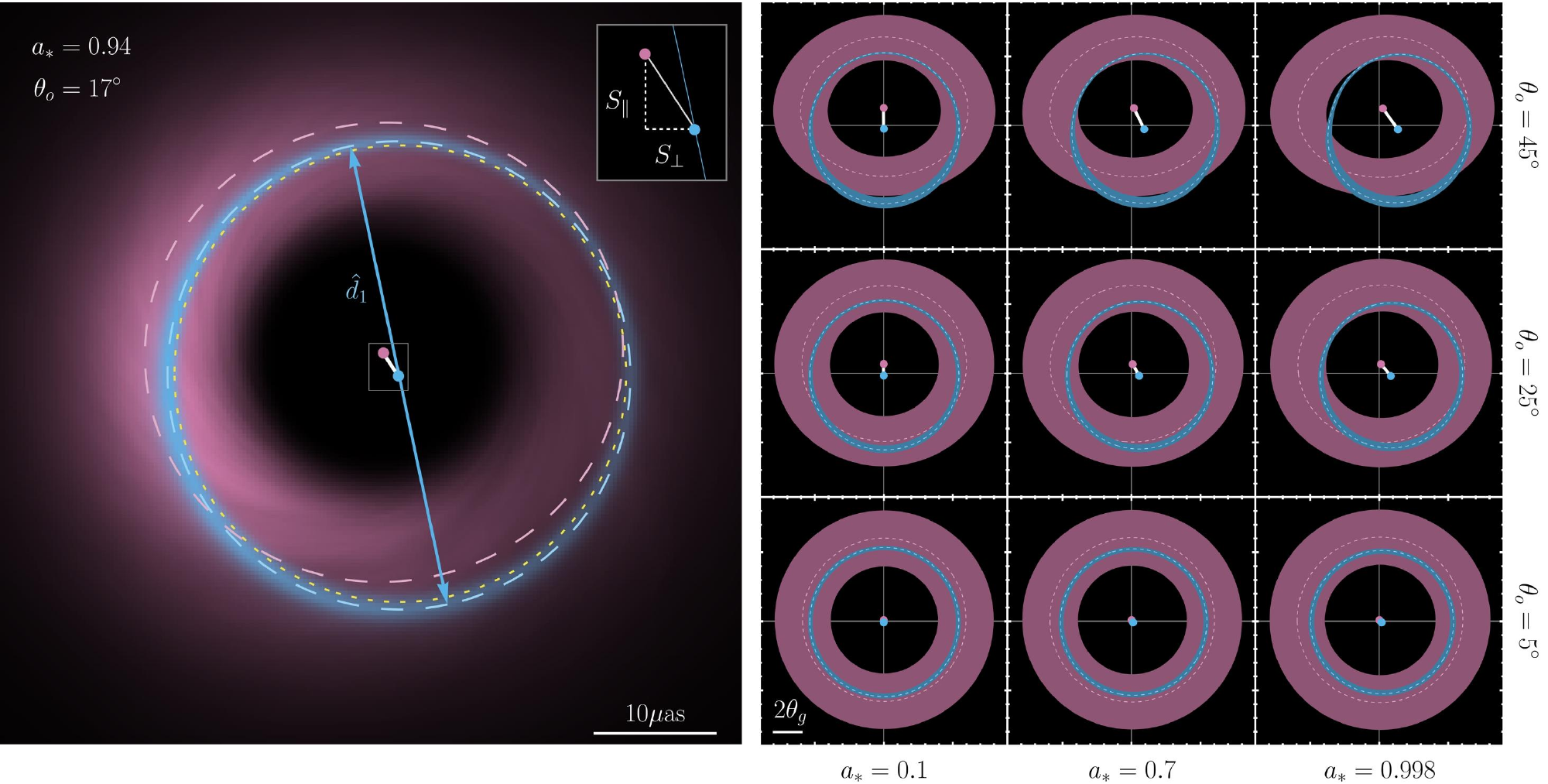}
    \includegraphics[width=.67\linewidth]{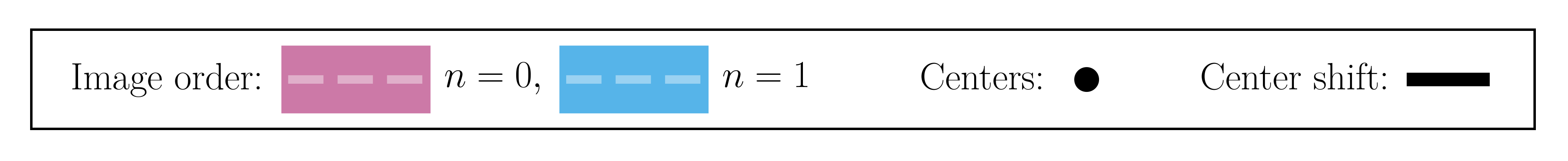}
    \caption{
    {Horizon-scale images of the direct (${n}~{=}~{0}$; in pink) and secondary (${n}~{=}~{1}$; in blue) sub-images of emitting accretion disks around spinning black holes whose projected spin axes are oriented vertically. The sub-image centers (correspondingly colored points), given by the centroids of sub-image contours (correspondingly colored dashed contours), are used to define the relative location (solid white lines) of the direct and secondary sub-images. 
    {\bf Left:} Prograde MAD GRMHD simulation with $a_*=0.94$, $\theta_o=17^\circ$, and $R_{\rm high}=R_{\rm low}=1$. 
    Each sub-image is given by the ellipse whose Gaussian convolution best fits the brightness profile. 
    We include an inset illustrating the parallel and transverse center shifts ($S_\parallel$ and $S_\perp$), the maximum diameter of the ${n}~{=}~{1}$ sub-image contour ($\hat{d}_1$; marked with blue double arrow), and the critical curve (dotted yellow line).
    {\bf Right:} Geometric equatorial accretion disk model spanning radii ${r_s}~{\in}~{[3r_g,7r_g]}$, shown for varying black-hole spins (columns) and observer inclinations (rows). 
    Each sub-image contour is the projection of the disk's average radius (${r_s}~{=}~{5r_g}$).}
    }
    \label{fig:OnSkyDisks}
\end{figure*}

As defined, the transverse and parallel shifts scale linearly with the mass-to-distance ratio of the black hole.
It is advantageous to define a common metric that can be applied to any black-hole image. 
A simple approach would be to divide our shifts by the black hole's mass, but various measurement techniques yield varying mass estimates for \m.
This approach would then suffer from systematic errors that would depend on a given choice of mass measurement. We note, for example, mass estimates from \cite{M87_mass_gas_Walsh}, \cite{M87_mass_stellar_Liepold}, \cite{M87_mass_stellar_Simon}, and \cite{m87_2018_I} which can vary by nearly an order of magnitude{.}
We thus propose to use a self-consistent normalization in terms of black hole image features.
Some natural candidates for normalization are the sizes of the ${n}~{=}~{0}$ and ${n}~{=}~{1}$ sub-image rings. 
As the size of the secondary sub-image is less sensitive to changes in astrophysics and black-hole spin than the direct sub-image, we choose the {maximal} diameter of the ${n}~{=}~{1}$ sub-image, $\hat d_1$, as our normalization.
Thus, we study the normalized parallel and normalized transverse shifts: 
\begin{align}
    \label{eq:CenterShifts}
    S_{\parallel}=\frac{|y_0-y_1|}{\hat d_1},\quad
    S_{\perp}=\frac{|x_0-x_1|}{\hat d_1}.
\end{align}

As images of accretion disks exhibit extended emission of varying brightness, the geometric notion of the normalized shifts is model-dependent.
In~\autoref{fig:OnSkyDisks}, we depict the center shifts of a {general-relativistic magnetohydrodynamics (GRMHD) simulation in the MAD state (details in \autoref{sec:GRMHD}) and for an analytic equatorial disk model.}
For the GRMHD case (left panel), we define the center shifts with respect to Gaussian-blurred {ellipses} that we fit directly to the image. For the geometric equatorial disk case (right panel), we define the center shifts with respect to the average radial extents of the disk's sub-images.
{At low and moderate inclinations, the normalized parallel shifts are relatively insensitive to spin, while the normalized transverse shifts are primarily driven by the spin.
This is the central physical effect that explains why relative photon-ring astrometry can constrain the black-hole spin, and is visualized in the right panel of~\autoref{fig:OnSkyDisks}.}

In this work, we aim to link the center shift to black-hole system parameters (like spin, inclination, and mass) that are independent of astrophysics, so it is advantageous to define the sub-image centers of the image-plane brightness profile using models that mitigate differences from brightness variations. 
More precisely, models that assign a geometric interpretation to brightness variations can be used to define a sub-image contour that serves as a proxy for the average shape, size, and location of the accretion-disk image on the image plane. 
Thus, a suitable choice {is} the geometric models employed by the EHTC to extract geometric information from images of \m \citep{M872018_I}\footnote{In particular, the EHTC geometric models capture the relatively fixed size of the \m direct image, while modeling the changing brightness distribution across years of observation.}. 
We note explicitly that the commonly used center-of-light centroid, the intensity-weighted mean position on the image plane, is not ideal for our analysis. The ${n}~{=}~{0}$ to ${n}~{=}~{1}$ center-of-light centroid displacement for \m was explored in \cite{GutierrezLara_interferometric_phase} where it was shown to be similarly sensitive to astrophysical parameters, such as plasma speed, as it is to spin.

\section{Center shifts in a semi-analytic equatorial ring model}
\label{sec:ERM}

We take inspiration from GRMHD simulations of \m with MADs to develop a semi-analytic model of the center shifts in time-averaged images of black hole accretion flows.
MAD simulations typically feature mm-wavelength emission dominated by material at low scale heights and located within {a few gravitational radii ($\lesssim~{5r_g}$) \citep{Dexter_emission_origin,EHTC_M87_V,Emami_Emissivities_2023}.}
We will assume that the sub-image contours of both the direct and secondary sub-images can be approximated by the ${n}~{=}~{0}$ and ${n}~{=}~{1}$ projections of a single equatorial source radius ${r_s/r_g}~{\leq}~{10}$ {where we adopt this outer radius bound to be conservative and to account for the possibility of extended structure}.
We will refer to this model as the ``equatorial ring model''. 

On each sub-image, an infinitesimally thin equatorial (${\theta_s}~{=}~{\pi/2}$) source ring is mapped to a closed contour.
(See, for example, in the right panel of \autoref{fig:OnSkyDisks}, the direct ${n}~{=}~{0}$ and secondary ${n}~{=}~{1}$ sub-images of equatorial rings at ${r_s/r_g}~{=}~{3,\ 5,}$ and $7$ correspond to the inner edge, dashed midline, and outer edge of the shaded regions{.}) 
Without loss of generality, we can specialize our study to viewing inclinations ${\theta_o}~{\leq}~{90^\circ}$ because mapping ${\theta_o}~{\mapsto}~{\pi}~{-}~{\theta_o}$ flips the projected contour of an axisymmetric equatorial ring about the $x$-axis on the image-plane.

We define the sub-image center for the projected contour of an equatorial ring to be its arc-length-weighted centroid. 
With the center of the sub-image contour defined, we can construct an image-plane diameter at each image-plane angle ($\rm arctan(y/x)$) as the length of the chord (i.e., straight line segment which joins two points on the contour) passing through the center. 
The equatorial ring model is expected to degenerate at large observer inclinations as the ring's sub-images become more irregular (i.e., concave).
In addition, we expect the scale height of the emission to become more important when black-hole-accretion systems are viewed at higher inclinations. (See, for example, \cite{jukebox}, where the authors find that equatorial emission models provide worse fits to black-hole images at higher inclinations{.})
As such, we will restrict our study to examining the center shifts for observers at inclinations ${\theta_o}~{\leq}~{45^\circ}$. 
In summary, we examine the shifts on the domain of interest: 
\[(0\leq{a_*}\leq1){,}~(0^\circ\leq{\theta_o}\leq45^\circ){,}~(2\leq{r_s/r_g}\leq10).\]

The sub-image centers of the equatorial ring model allow us to elucidate the typical behavior of the relative locations of ${n}~{=}~{0}$ and ${n}~{=}~{1}$ sub-images (see the right panel of \autoref{fig:OnSkyDisks}). 
We find that increasing the observer's inclination tends to increase the {parallel shift} ($|{y_0}~{-}~{y_1}|$), growing as the ${n}~{=}~{0}$ center translates upward on the image plane. 
The ${n}~{=}~{1}$ center translates downward, but to a much smaller degree. 
For non-zero viewing inclinations, increasing {the} spin tends to increase the {transverse shift} ($|{x_0}~{-}~{x_1}|$) as the ${n}~{=}~{1}$ center translates rightward on the image plane, with the ${n}~{=}~{0}$ center also translating rightward to a lesser degree. 
We also find that the magnitude of the spin-driven shift is enhanced at larger inclinations~\footnote{These center shift behaviors appear to persist for alternative center definitions, e.g., the area centroid of the image-plane region enclosed by the ${n}~{=}~{1}$ sub-image contour.}. 
{See \autoref{app:EqRingImages} for details on image-plane projections of equatorial rings.}

\begin{figure}
    \centering
    \includegraphics[width=\linewidth]{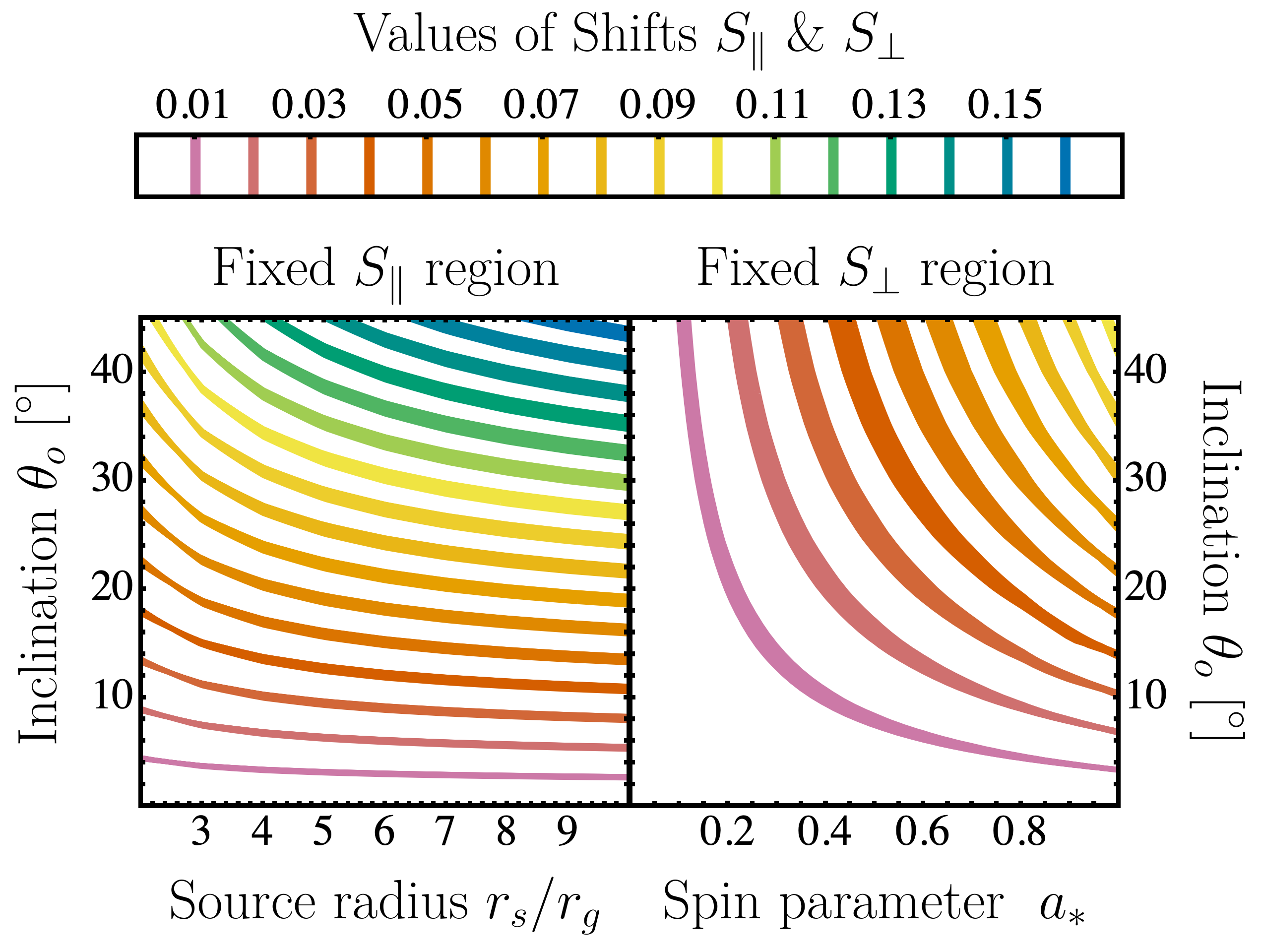}
    \caption{Feasibility regions in which fixed shift values {$\cu{0.01,0.02,\dots,0.16}$ (colored region)} can be achieved. 
    {The} normalized parallel shift $S_\parallel$ {is} marginalized over all spins ${0}~{\leq}~{a_*}~{\leq}~{1}$ (left){;} the normalized transverse shift $S_\perp$ {is} marginalized over source radii ${2}~{\leq}~{r_s/r_g}~{\leq}~{10}$ (right).}
    \label{fig:Feasibility}
\end{figure}
The normalized center shifts, \autoref{eq:CenterShifts}, naturally reflect the behavior of their unnormalized counterparts while additionally exhibiting useful characteristics beyond their mass-to-distance independence~\footnote{{Choosing alternative ${n}~{=}~{1}$ diameters (e.g., minimal or average chord diameter, or minimal, maximal, or average projected diameter) to normalize the {shifts} are expected to yield similar results.}}.
We find that $S_{\parallel}$ to be relatively insensitive to the black-hole spin, while $S_{\perp}$ to be relatively insensitive to the emission source radius.
These insensitivities are illustrated in \autoref{fig:Feasibility}, where we plot feasibility regions for fixed values of $S_{\parallel}$ (marginalized over all spins) and $S_{\perp}$ (marginalized over all source radii ${2}~{\leq}~{r_s/r_g}~{\leq}~{10}$). 
The left panel shows that an observed value of $S_{\parallel}$ bounds all allowed combinations of inclination and source radius, imposing an upper and lower bounds on $\theta_o$ for each source radius.
The right panel shows that an observed value of $S_{\perp}$ bounds all allowed combinations of inclination and spin, imposing lower bounds on both inclination and spin.

\begin{figure}
    \centering
    \includegraphics[width=\linewidth]{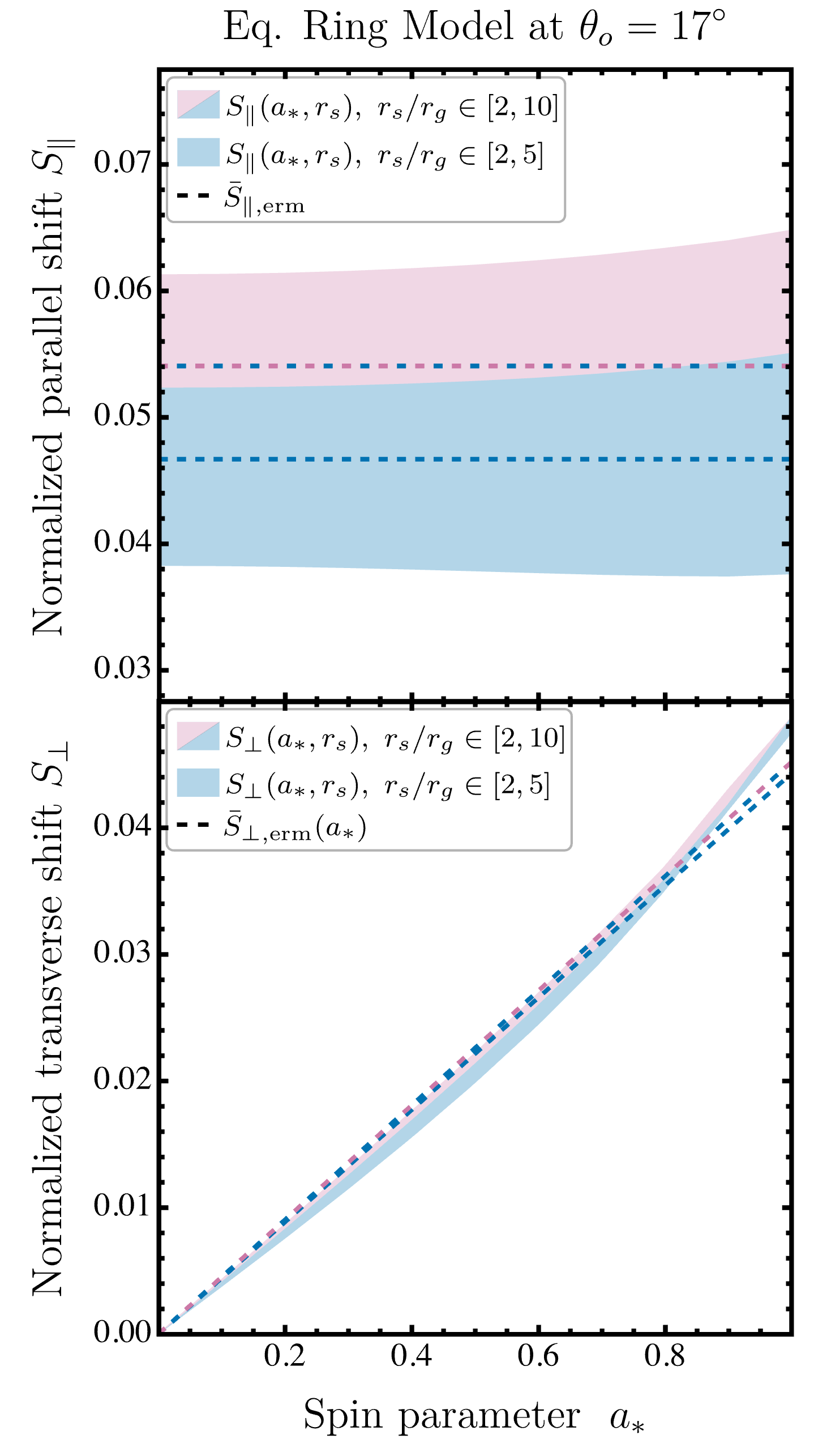}
    \caption{{Normalized parallel (top) and transverse (bottom) shift regions of equatorial rings with ${r_s/r_g}~{\in}~{\br{2,10}}$ (combined pink and blue) and with ${r_s/r_g}~{\in}~{\br{2,5}}$ (blue only), viewed at ${\theta_o}~{=}~{17^\circ}$, as a function of spin alongside the corresponding best-fit mean values (\autoref{eq:vERMFit} and \autoref{eq:hERMFit}; dashed lines). 
    }}
    \label{fig:FixedInclination}
\end{figure}

An important implication of the sensitivity of $S_\parallel$ and $S_\perp$ to the observer inclination is that a constraint on $\theta_o$ places bounds on both quantities. 
In particular, a known $\theta_o$ can be combined with a measurement of $S_\perp$ to constrain the spin parameter $a_*$. 
We examine $S_\parallel$ and $S_\perp$ for an \m-like viewing inclination of ${\theta_o}~{=}~{17^\circ}$.
{In \autoref{fig:FixedInclination}, we depict the range of shift values that result when the source radius is bounded by ${2}~{\leq}~{r_s/r_g}~{\leq}~{10}$ or the more restrictive range ${2}~{\leq}~{r_s/r_g}~{\leq}~{5}$.}

{More realistically, the observed center shifts are sourced by emission from an ensemble of radii in the spacetime. 
With the equatorial ring model, we model the expected shifts as a distribution resulting from random draws of spin and source radius pairs.
We assume both the spin and source radius are drawn from uniform priors within our range of interest. 
We approximate the movements of such a distribution of center shifts as the least squares means $\left(\bar S_\perp \text{ and }\bar S_\parallel\right)$ and sample standard deviations $\left(\sigma_{\perp} \text{ and } \sigma_{\parallel}\right)$. Motivated by the shape of the feasibility bands in the equatorial ring model, we assume a spin-independent model for $\bar{S}_{\parallel}$, and a $\bar{S}_\perp$ model that is proportional to the spin. Performing a continuous least squares regression yields the approximators:}
\begin{subequations}
\label{eq:vERMFit}
\begin{align}
    \hat{S}_{\parallel}=&{\bar{S}_{\parallel,\rm{erm}}}\pm {\sigma_{\parallel,\rm{erm}}},\\
    =&
    \begin{cases}
    (5.4\pm 0.7)\cdot10^{-2}, \quad 2\leq r_s/r_g\leq 10\\
    (4.7\pm 0.5)\cdot10^{-2}, \quad 2\leq r_s/r_g\leq 5
    \end{cases}{,}
\end{align}
\end{subequations}
\begin{subequations}
\label{eq:hERMFit}
\begin{align}
    \hat{S}_{\perp}= &{{\bar{S}_{\perp,\rm erm}} (a_*)}\pm {\sigma_{\perp,\rm erm}},\\
    =&
    \begin{cases}
    (4.5a_*\pm0.1)\cdot10^{-2}, \quad 2\leq r_s/r_g\leq 10\\
    (4.4a_*\pm 0.1)\cdot10^{-2}, \quad 2\leq r_s/r_g\leq 5
    \end{cases}.
\end{align}
\end{subequations}
(Note {that} we have added the additional subscript ``erm'' in anticipation of comparing the best-fit shift values derived from the equatorial ring model to those we will derive from GRMHD simulations in the next section.) The {best-fit} mean shift values, {${\bar{S}_{\parallel}}$} and {${\bar{S}_{\perp}}$}, are also shown in \autoref{fig:FixedInclination}.

Comparing the best-fit values for the conservative source radius range (${2}~{\leq}~{r_s/r_g}~{\leq}~{10}$) and the more restrictive range (${2}~{\leq}~{r_s/r_g}~{\leq}~{5}$), we find that smaller source radii result in smaller values of $S_{\parallel}$.
Narrowing the range of source radii also slightly narrows the spread of $S_\perp$. 
The narrower radii range results in a shallower slope for the linear fit of $S_{\perp}$ as a function of spin.
However, we note that the difference between the slopes for the conservative and restricted ranges of source radii is very slight, consistent with the expected behavior that $S_\parallel$ is more sensitive to source radii than $S_\perp$. 

We reiterate that, in defining the equatorial ring model we have assumed that both the ${n}~{=}~{0}$ and ${n}~{=}~{1}$ sub-images of near-horizon black-hole accretion can be approximated by emission from a single source radius in the equatorial plane.
(\citet{bhex_science} has also used a similar equatorial ring model, neglecting the effects of the small but non-zero inclination, to argue that comparing the sizes of the direct and secondary images may be used to constrain the black-hole spin.)
The suitability of our assumption has not been tested, but the shifts derived from the equatorial ring model at ${\theta_o}~{=}~{17^\circ}$ are surprisingly robust even if the assumption is relaxed. 
Changes in the radial domain of interest induce a negligible change in the fitted means in \autoref{eq:vERMFit} (${\approx}~{0.2\%}$ change) and \autoref{eq:hERMFit} (${\approx}~{0.3\%}$ change). 
Thus{,} the center shifts are relatively agnostic to the precise source radii at which the ${n}~{=}~{0}$ and ${n}~{=}~{1}$ emission originates, provided the vast majority of the emission is sourced within the first few gravitational radii.

Next, we compare these center shifts from the equatorial ring model to those present in GRMHD simulations.

\section{Center shifts in GRMHD simulations}
\label{sec:GRMHD}

In this section, we study the center shifts in horizon-scale images of realistic accretion flows.
We examine the normalized parallel and normalized transverse shifts present in 
{synthetic time-averaged images of \m produced for the analysis performed in \cite{Palumbo_Wong_2022}~\footnote{{In detail, the GRMHD simulations were performed as part of the \texttt{PATOKA} pipeline~\citep{patoka}, which produced the v1 Illinois library that contributed to the EHT M87 library used for analysis of the first EHT observation~\citep{EHTC_M87_V,EHTC_M87_VI}, but were ray traced anew to create the \cite{Palumbo_Wong_2022} image library.}}}.
As previously mentioned, there are two classes of accretion states that typically arise in GRMHD simulations: {MAD} and {SANE}.
We focus our attention on MAD simulations, as they are {the preferred accretion state of \m inferred from the observed jet power and horizon-scale polarization structure~\citep{EHTC_M87_V,EHTC_M87_VIII}.}
We examine simulations with spin values ${a_*}~{\in}~{\cu{0,\pm0.5,\pm0.94}}$ and {ion-electron temperature ratio parameters ${R_{\rm high}}~{\in}~{\cu{1,10,20,40,80,160}}$ and fixed ${R_{\rm low}}~{=}~{1}$}.
Following the convention of the EHTC, we now imbue the black-hole spin parameter with a sign (${{\rm sign}(a_*)}~{=}~{\pm}$) to distinguish between accretion disks that rotate prograde/retrograde at large gravitational radii compared to the rotational sense of the black hole. 

The GRMHD images under consideration here are ray traced at 230 GHz, a frequency {at which} the accretion disk is optically thin \citep{EHTC_M87_V}. In general, we expect the relative contribution of the ${n}~{=}~{1}$ image to increase with frequency as the optical depth of the accretion disk decreases 
\citep{Palumbo_2024, Palumbo_2025,Desire_multifrequency_models}. Moreover, we expect that GRMHD images of insufficient resolution may undersample the ${n}~{=}~{1}$ ring, as discussed in detail in \citet{Gelles_2021} and \citet{Desire_multifrequency_models}. Our images contain 480 pixels in each dimension across a 160 $\mu$as field of view, with an angular gravitational radius of 3.83 $\mu$as. The resulting pixel size of $\sim0.087$ angular gravitational radii is convergent for the $n=1$ ring, which is sufficient for our study. However, the ${n}~{=}~{2}$ and higher rings are underrepresented in the image.

In contrast to the equatorial ring model, the center shifts we measure from a simulated image depend on its brightness distribution rather than the geometric shapes on the image plane. 
As such, we measure the center shifts by fitting a model of the brightness distribution to GRMHD images that permits a geometric interpretation for the brightness distribution, allowing us to characterize the general sizes and locations of the direct and secondary disk sub-images. 

\begin{table*}
\centering
\begin{tabular}{llrr}
\hline
Parameter& Description & Lower bound & Upper bound \\
\hline
$d_{{n}}$& Largest diameter of the {$n$}-th mF-ring ($\mu\rm{as}$)      & 16.0    & 100.0 \\
$W_{{n}}$ & FWHM of the blurring kernel from the {$n$}-th mF-ring ($\mu\rm{as}$)  & 0.5    & 80.0 \\
rot.${}_{{n}}$ & Angle of the {$n$}-th mF-ring's major axis (counter-clockwise from north) (rads.) & $-\pi$ & $\pi$ \\
$\alpha_{1,{n}}$& First real component of brightness mode in the {$n$}-th mF-ring & -1.0   & 1.0 \\
$\beta_{1,{n}}$ & First imaginary component of brightness mode in the {$n$}-th mF-ring & -1.0   & 1.0 \\
$\alpha_{2,{n}}$ & Second real component of brightness mode in the {$n$}-th mF-ring& -1.0   & 1.0 \\
$\beta_{2,{n}}$& Second imaginary component of brightness mode in the {$n$}-th mF-ring & -1.0   & 1.0 \\
ell.${}_{{n}}$ & Ratio of smallest diameter to largest diameter of the {$n$}-th mF-ring     & 0.0    & 1.0 \\
floor${}_{{n}}$ & Brightness ratio between mF-ring component and floor in the {$n$}-th mF-ring & 0.0    & 1.0 \\
$x_{{n}}$ & Horizontal position of the {$n$}-th mF-ring  ($\mu\rm{as}$)  & -10.0  & 10.0 \\
$y_{{n}}$ & Vertical position of the {$n$}-th mF-ring   ($\mu\rm{as}$)      & -10.0  & 10.0 \\
flux rat.& Flux ratio between the {${n}~{=}~{0}$ and ${n}~{=}~{1}$ mF-rings} & 0 & 1
\end{tabular}
\caption{Table of parameters and their prior ranges used to fit pairs of mF-ring models to the time-averaged images of GRMHD simulations in  \autoref{sec:GRMHD} and \autoref{app:SANE}. {The subscript $n$ labels the specific mF-ring. We take $W_0\geq W_1$ to impose the association of the respective mF-rings with the direct and indirect accretion disk image components of the brightness profile.}}\label{tab:mring_params}
\end{table*}

In particular, we fit the accretion-disk image-plane brightness profile with a pair of {${m}~{=}~{2}$} mF-rings. 
The {${m}~{=}~{2}$} mF-ring model approximates a brightness distribution by convolving a thin ellipse, with {two} Fourier brightness modes distributed along it {and a flat brightness floor interior to it}, with a Gaussian \citep{M872018_I}. 
({This particular} mF-ring model was used in the EHTC's analysis of the 2018 observations of the total intensity of \m; for more details, see the appendix of \cite{M872018_I}.)
After fitting the pair of mF-rings to the image-plane brightness distribution, we use the de-biased best-fit mF-ring model ellipses as the sub-image contours from which we derive the centers and diameters. 
In detail, the MAD simulation center shifts are found through the following procedure:
\begin{enumerate}
    \item We produce time-averaged images of the combined ${n}~{=}~{0}$ and ${n}~{=}~{1}$ emission from 200 snapshots of each GRMHD simulation.
    Each snapshot image is separated by 25 gravitational times (${t_g}~{=}~{GM/c^3}$).
    These images are drawn from the library in \cite{Palumbo_Wong_2022}. 
    \item We use the pair of mF-ring models to perform geometric model fits to each image, using the \texttt{VIDA} template matching algorithm \citep{VIDA} to maximize the normalized cross-{correlation} between the mF-ring models and the MAD images.
    The fits are performed over the prior range described in \autoref{tab:mring_params}~\footnote{{We employed this mF-ring model for fitting the image-plane brightness profiles for consistency with \cite{M872018_I}. However, we note our fits preferred negligibly small floors in both the ${n}~{=}~{0}$ and ${n}~{=}~{1}$ mF-rings. As such, simple m-rings (lacking brightness floors interior to the ellipses) would yield comparable results.}}. 
    \item We use the best-fit parameters from our fit to extract the ${n}~{=}~{0}$ sub-image center $(x_0,y_0)$, the ${n}~{=}~{1}$ sub-image center $(x_1,y_1)$, and the maximum diameter of the ${n}~{=}~{1}$ ring $d_1$. 
    Since the convolution of the mF-ring with the Gaussian blur biases $d_1$ away from the peak of the brightness distribution, we use the FWHM ($W_{{1}}$) of the convolved Gaussian to report the de-biased diameter \citep{EHTC_SgrA_IV}
    \begin{equation}
        \hat d_1=d_1-\dfrac{W_1^2}{4\log(2)d_1}.
    \end{equation}
    \item We use these quantities to calculate the center shifts via \autoref{eq:CenterShifts}.
\end{enumerate}
The left panel of \autoref{fig:OnSkyDisks} shows one such image of a simulation (${a_*}~{=}~{0.94}$, ${R_{\rm{high}}}~{=}~{1}$) with the ${n}~{=}~{0}$ and ${n}~{=}~{1}$ de-biased ellipses, the ${n}~{=}~{0}$ and ${n}~{=}~{1}$ centers, and the ${n}~{=}~{1}$ maximal de-biased diameter derived using the mF-ring models.

\begin{figure*}
    \centering 
    {\Large MAD Simulations at $\theta_o=17^\circ$}\\
    \includegraphics[width=.497\linewidth]{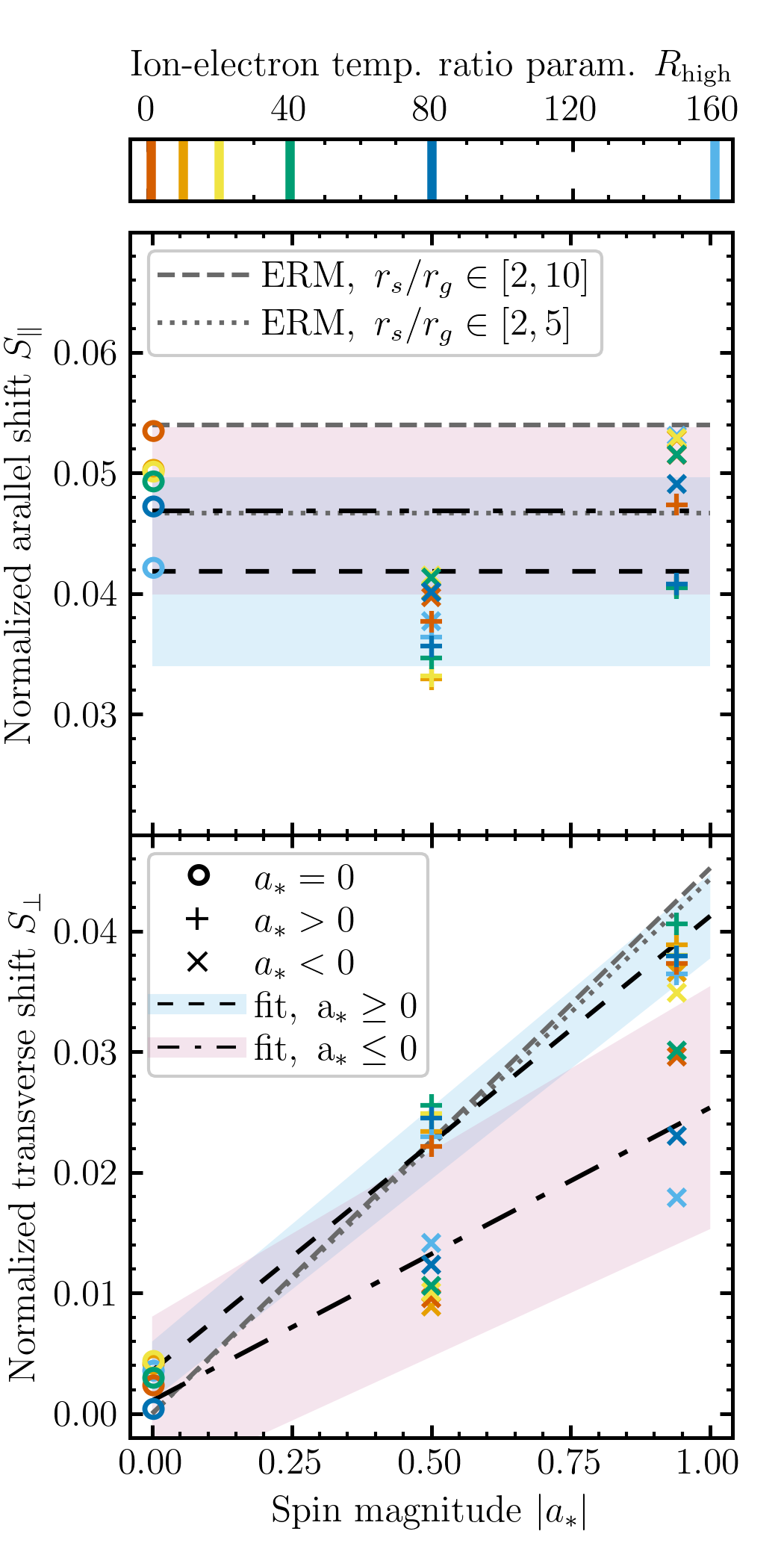}
    \includegraphics[width=.497\linewidth]{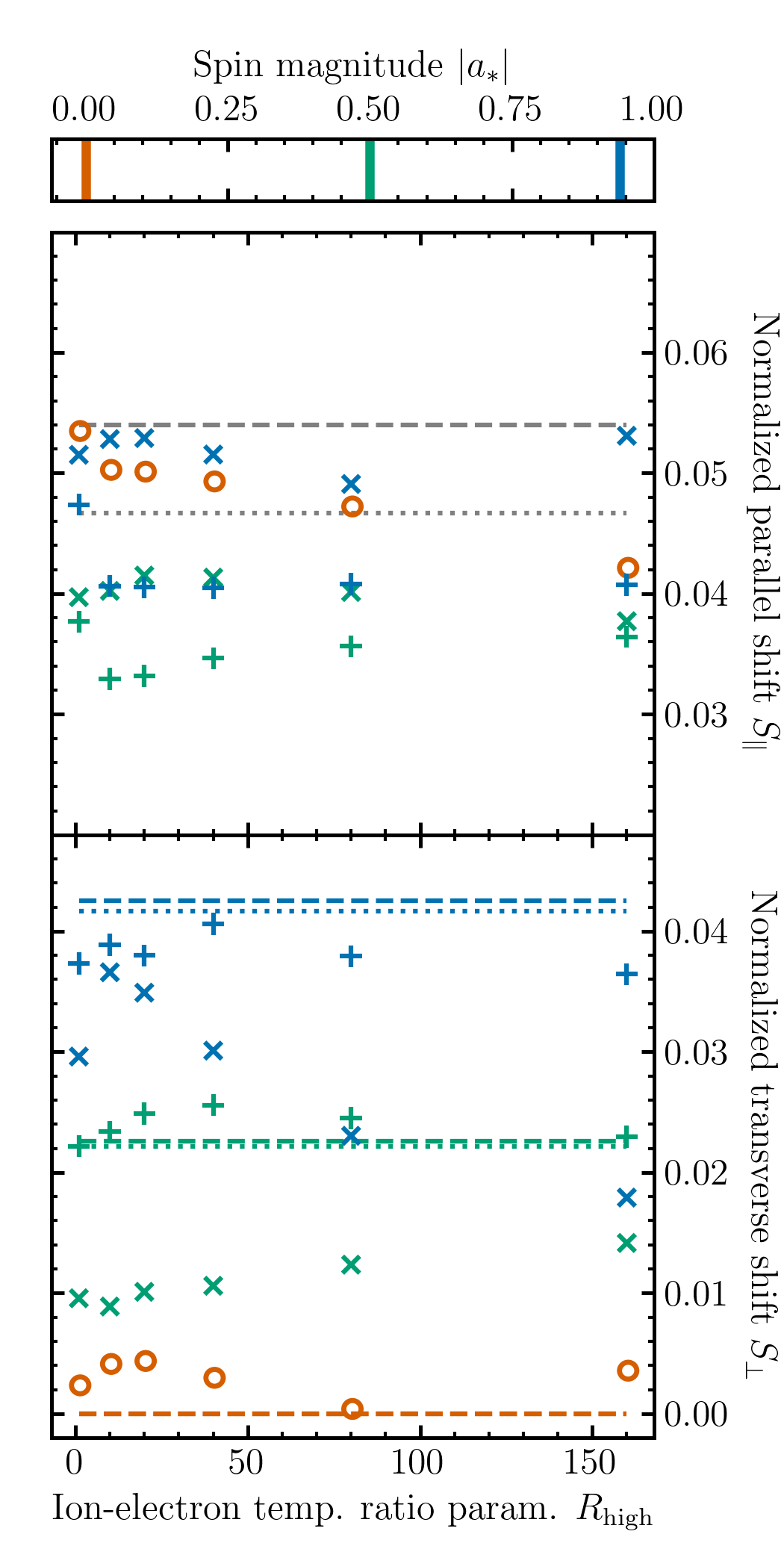}
    \caption{
    {
    Normalized parallel (top rows) and transverse (bottom rows) shifts in images of MAD GRMHD M87* simulations, viewed at ${\theta_o}~{=}~{17^\circ}$, as a function of spin (left) and $R_{\mathrm{high}}$ (right). 
    Marker style denotes accretion-flow rotation sense.
    Marker color denotes $R_{\mathrm{high}}$ on the left and spin on the right. 
    Best fits of GRMHD models are shown as light blue bands for prograde flows ($\hat{S}_{\parallel, +}$ \autoref{eq:vProFit} and $\hat{S}_{\perp,+}$ \autoref{eq:hProFit}) and pink bands for retrograde flows ($\hat{S}_{\parallel, -}$ \autoref{eq:vRetFit} and $\hat{S}_{\perp,-}$ \autoref{eq:hRetFit}).
    Best-fit mean values from the equatorial ring model (${\bar{S}_{\parallel,\rm erm}}$ \autoref{eq:vERMFit} and ${\bar{S}_{\perp,\rm erm}}$  \autoref{eq:hERMFit}) are shown as dashed and dotted lines in gray (top left, top right, bottom left) and spin-dependent colors (bottom right). }
    }
    \label{fig:MAD_GRMHD}
\end{figure*}

We summarize the fitted center-shift data in \autoref{fig:MAD_GRMHD}.
We observe that 1) $S_\parallel$ and $S_\perp$ exhibit no clear dependence on $R_{\rm high}$, 2) $S_\parallel$ exhibits no clear dependence on $a_*$, and 3) $S_\perp$ increases with $a_*$.
The spin-associated trends seen here are similar to those found in the equatorial ring model.

As the shift measured in images of simulated accretion flows {is} a property of the brightness distributions, we expect our choice of image model (the mF-ring in this case) and additional accretion effects---which can cause angular and radial dependence in emission---may cause deviations from our predicted shift behavior.
We assume these additional-error-introducing effects are independent of spin{.}

{We approximate the moments of the underlying distribution of center shift as least squares means $\left(\bar S_\perp \text{ and }\bar S_\parallel\right)$ and the standard deviations $\left(\sigma_{{\perp}} \text{ and } \sigma_{{\parallel}}\right)$ of our sample of center shift measurements from \autoref{fig:MAD_GRMHD}. Again, we assume a spin-independent model for the parallel shift $\bar{S}_\parallel$. However, we adopt a general linear model for the transverse shift $\bar{S}_\perp=\bar{m}|a_*|+\bar{b}$.}
{This defines the approximators,}
\begin{align}
    \hat S_{{{\parallel}}}&=(\bar{S}_{{\parallel}}\pm\sigma_{{\bar S_{\parallel}}})\pm{\sigma_{\parallel}}, \label{eq:ParallelShiftApproximator}\\
    \hat S_{{\perp}} (a_*)&=({\bar{m}}\pm\sigma_{{{\bar m}}})\,|a_*| +({\bar{b}}\pm\sigma_{{{\bar b}}})\pm{\sigma_{\perp}}, \label{eq:TransverseShiftApproximator}
\end{align}
where $\sigma_{{\bar S_{\parallel}}}$, $\sigma_{{\bar m}}$, and $\sigma_{{\bar b}}$ are the standard errors from our least squares fit.
{We will consider separately the} cases where we can and cannot discern between the accretion disk's rotation senses. {When explicitly necessary, we will include an} additional subscript {to denote} the cases {in which} the accretion-flow rotation sense is prograde (`$+$'), retrograde (`$-$'), or indistinguishable (`$\pm$'). 

{We obtain the} normalized parallel shifts
\begin{align}
        \hat S_{\parallel,+}&=
    {\br{(4.2\pm0.1)\pm 0.6}\cdot10^{-2}}, \label{eq:vProFit}\\
        \hat S_{\parallel,-}&=
    {\br{(4.6\pm0.1)\pm 0.6}\cdot10^{-2}},\label{eq:vRetFit}
\end{align} 
for the prograde and retrograde cases, shown in the top-left panel of \autoref{fig:MAD_GRMHD}, and
\begin{align}
    \hat S_{\parallel,\pm}&={
    \br{(4.3\pm0.1)\pm 0.7}\cdot10^{-2}},
    \label{eq:vBothFit}
\end{align}
for the indistinguishable case~{\footnote{\label{ftnt:wls}{Note that here we perform ordinary least squares regression on the sample of measurement shifts, which amount to shift derived when assuming approximately uniform priors on spin and on ${\rm log}R_{\rm high}$. Such $R_{\rm high}$ sampling is common in EHT analyses (e.g., \cite{EHTC_M87_V,EHTC_M87_VIII}). For fits assuming an approximately uniform prior on $R_{\rm high}$, see \autoref{app:Weights}.}}.
We note that, for the parallel shifts in the GRHMD simulations, our best-fit prograde/retrograde mean is lower than/comparable to those of the equatorial ring model \autoref{eq:vERMFit}.
The spread of the GRMHD fits is comparable to the equatorial ring model.}
Comparing the prograde and retrograde cases, we find that the retrograde {case} has a larger mean and spread than the prograde {case}. 

We also obtain the prograde and retrograde normalized transverse shifts:
\begin{align}
    \hat S_{\perp,+}&= {\br{(3.8\pm0.1)|a_*| +(0.4\pm0.2)\pm 0.2}\cdot10^{-2}},
    \label{eq:hProFit}\\
    \hat S_{\perp,-}&={\br{(2.7\pm0.3)|a_*| +(0.1\pm0.2)\pm0.5}\cdot10^{-2}},
    \label{eq:hRetFit}
\end{align}
which we plot in the bottom-left panel of \autoref{fig:MAD_GRMHD}.
Additionally, we find
\begin{align} 
    \hat S_{\perp,\pm}&={\br{(3.3\pm0.3)|a_*| +(0.2\pm0.2)\pm0.6}\cdot10^{-2}},
    \label{eq:hBothFit}
\end{align}
for the indistinguishable case~${}^{\ref{ftnt:wls}}$.
Compared to the predicted behavior of the equatorial ring model of \autoref{eq:hERMFit}, the additional effects in the GRMHD simulations bias both the intercept and the slope.
{In all fits, we find a larger spread and a shallower slope than the equatorial ring model, as well as a non-zero $y$-intercept.}
Furthermore, the retrograde case has {a} much shallower slope than the prograde case, indicating a greater departure from the equatorial ring model.

{The differences between the simulation and equatorial ring model may have physical origins (e.g., accretion-flow and lensing properties) that are not captured in our highly idealized model.}
{In the equatorial ring model, the normalized transverse shift as a function of spin is steeper than in the simulations. This aligns with the fact that MAD flows do exhibit some non-equatorial emission, which the equatorial ring model neglects \citep{Dexter_emission_origin,EHTC_M87_V,Emami_Emissivities_2023}.}
Heuristically, this {effect} can be understood by considering images of fixed-radius rings that are lifted above the equatorial plane, thereby bringing them slightly closer to the observer.
{And, at fixed radius,} frame dragging decreases away from the equatorial plane. 
Thus, a non-equatorial ring should exhibit less normalized transverse shift on the image plane. 

Comparing the center shifts across the prograde and retrograde cases may hint at {the} key differences between {the two cases}. 
First, the larger mean and larger error of the normalized parallel shift in retrograde flows {is consistent with the fact} that, in the retrograde case, the radii contributing to the horizon-scale emission are on average farther from the black hole and span a broader range than in prograde flows {\citep{Emami_Emissivities_2023}}.
Second, the normalized transverse shift as a function of spin is shallower in retrograde flows than in prograde flows. 
This {is consistant with the fact} that retrograde flows exhibit less equatorially dominated emission than prograde flows {\citep{Emami_Emissivities_2023}}.

Finally, we note that while we have focused on MAD simulations, EHT observations disfavor but do not completely rule out the SANE accretion state for \m.
The equatorial ring model is expected to better describe the observed center shifts in images of {MADs} which tend to exhibit more disk-dominated (closer to the equatorial plane) emissivities than SANEs~{\citep{EHTC_M87_V,Emami_Emissivities_2023}}. 
In \autoref{app:SANE}, we calculate the center shift of SANE simulations and show its mismatch with the equatorial ring model.

\section{Discussion}
\label{sec:Discussion}

We have shown that a simple equatorial ring model produces a relative centroid shift between the ${n}~{=}~{0}$ and ${n}~{=}~{1}$ sub-images that is sensitive to emission radius, observer inclination, and black-hole spin.
In particular, we find that, for a black hole of known inclination, the magnitude of the normalized transverse shift $S_\perp$ is strongly indicative of spin.
Although our model is idealized, we show that $S_\perp$ can be robustly defined from the intensity distribution observed in radio images of black-hole accretion flows, despite the presence of potential errors arising from unmodeled astrophysical effects.
The relative astrometric displacement of the sub-image components could thus serve as a useful observable for spin measurements of \m in high-resolution black-hole imaging applications, such as the future BHEX space VLBI mission \citep{bhex_concept, bhex_instrument}.

Our results allow us to determine the astrometric requirements necessary for high-resolution black-hole imaging observations to constrain spin.
We find a relationship between the spin uncertainty ($\sigma_{{|a_*|}}$), the model-parameter uncertainties ($\sigma_{{\bar{m}}}$ and $\sigma_{{\bar{b}}}$), {standard deviation ($\sigma_\perp$),} and the center-shift measurement uncertainty ($\sigma_{\rm meas}$):
\begin{align}
    \label{eq:SpinMeasurementPrecision}
    \sigma^2_{{|a_*|}}= \frac{1}{\bar{m}^2}\left(a_*^2\sigma_{{\bar{m}}}^2 +\sigma_{{\bar{b}}}^2+{\sigma}^2_{{\perp}}+\sigma^2_{\rm meas}\right).
\end{align}
The {center-shift} measurement uncertainty is related to relative astrometric resolution {of} the observing instrument ($\sigma_{\rm r.a.}$) and the on-sky photon-ring diameter ($\hat{d}_1$) via $\sigma_{\rm meas}=\sigma_{\rm r.a.}/\hat{d}_1$. 

\begin{figure}
    \centering
    \includegraphics[width=\linewidth]{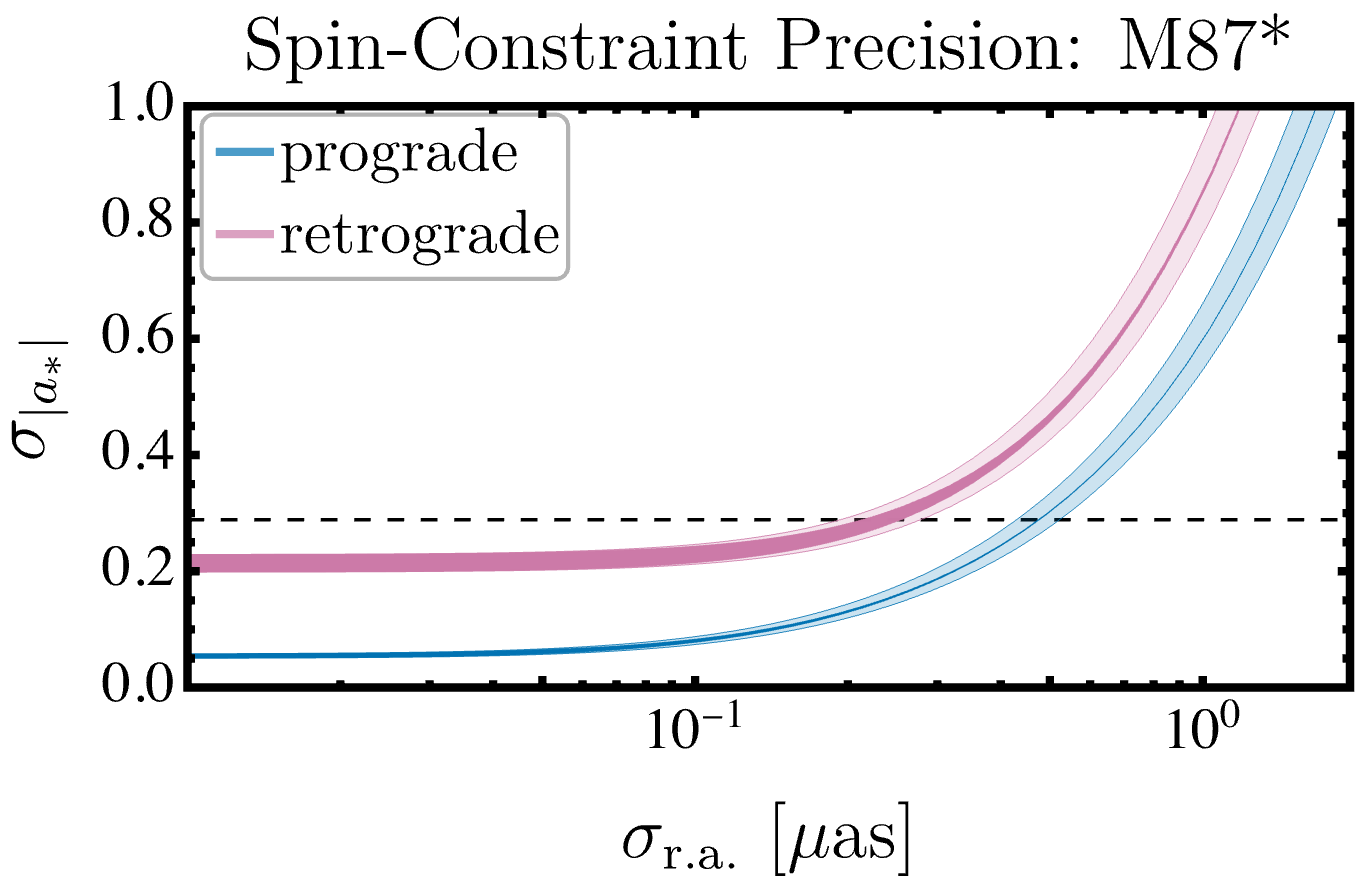}
    \caption{
    Spin-constraint precision $\sigma_{|a_*|}$ for \m with prograde (blue region) and retrograde (pink region) accretion flows as a function of relative astrometric error $\sigma_{\rm r.a.}$. In the darker regions, the ${n}~{=}~{1}$ ring diameter $\hat{d}_1$ is taken to be coincident with that of the ${n}~{=}~{0}$ ring and the spread in $\sigma_{{|a_*|}}$ accounts only for variation over all spin magnitudes ${0}~{\leq}~{|a_*|}~{\leq}~{1}$. In the lighter regions, the ${n}~{=}~{1}$ diameter is taken to lie anywhere within the ${n}~{=}~{0}$ ring. For comparison, the standard deviation corresponding to a flat distribution in spin is also shown (black-dashed line).
    }
    \label{fig:SpinConstrainPrecision}
\end{figure}

In typical accretion models, the photon ring lies on top of the direct image, which allows us to place a general bound on the size of the ${n}~{=}~{1}$ accretion-disk image using EHTC measurements.
The most recent \m measurements reported by \cite{M872018_I}, inferred from mF-ring model fits, give the direct-image diameter of ${\hat{d}_0}~{=}~{44.5(+0.69,-0.60)~\mu\rm{as}}$ and fractional width of ${\hat{f}_{\rm w}}~{=}~{0.19(+0.04,-0.05)}$, where we adopt the measurement from band 3 on April 21, 2018, since that dataset had the most stable imaging results.
Using this measurement, we present estimates of the spin-constraint precision of \m in the cases of prograde and retrograde accretion flow. 
In \autoref{fig:SpinConstrainPrecision}, we show the spin-constraint precision as a function of the relative astrometric resolution when the ${n}~{=}~{1}$ sub-image contour diameter is equal to that of the ${n}~{=}~{0}$ ring $\pa{{\hat{d}_1}~{=}~{\hat{d}_0}~{=}~{44.5\mu\rm{as}}}$, and when it is bounded by the inner and outer diameters of the ${n}~{=}~{0}$ ring $\pa{{\hat{d}_1}~{\in}~{\hat{d}_0 [1-\hat{f}_{\rm w}/2,1+\hat{f}_{\rm w}/2]} \text{ for } {\hat{f}_{\rm w}}~{=}~{0.19}}$.
We note that at large relative astrometric resolution, the spin-constraint precision is dominated by the ${n}~{=}~{1}$ sub-image diameter; whereas, at small relative astrometric resolution, it is dominated by the astrophysical uncertainty. 

Using these geometric estimates, we may relate the predicted spin uncertainty to the astrometric resolution of a VLBI array. Crucially, VLBI routinely recovers astrometric shifts much smaller than the beam for ``in-beam'' separations, where no absolute reference is needed. For example, \citet{Fomalont_1999} achieved sub-milliarcsecond relative astrometric resolution with the Very Long Baseline Array using in-beam referencing, improving on the nominal beam by more than an order of magnitude. More generally, given a sufficiently good model for the on-sky image, the astrometric resolution can improve on the nominal beam by a factor limited only by thermal noise, as described in Appendix 12.1.3 of \citet{TMS_2017}:
\begin{align}
    \sigma_{{\rm r.a.}} &\approx \frac{\sigma_\phi}{2 \pi \sqrt{N} B_\lambda}.
\end{align}
Here, $\sigma_\phi$ is the typical phase error on the baseline, $N$ is the number of observations, and $B_\lambda$ is the baseline length measured in wavelengths. For a mission approaching ${1/B_\lambda}~{\approx}~{6\;\mu\rm{as}}$ such as BHEX, repeated observations with a signal-to-noise ratio exceeding 5 can easily reach an astrometric resolution ${\sigma_{\rm r.a.}}~{\lesssim}~{0.1\;\mu\rm{as}}$. 

\autoref{eq:SpinMeasurementPrecision} imposes a condition on the relative astrometric resolution required to constrain the black-hole spin to within $\sigma_{{|a_*|}}$:
\begin{align}
    \sigma_{\rm r.a.}\leq \hat{d}_1\sqrt{\bar{m}^2\sigma_{{|a_*|}}^2-\pa{a_*^2\sigma_{{\bar{m}}}^2 +\sigma_{{\bar{b}}}^2+{\sigma_{\perp}}^2}}. \label{eq:RelAstro}
\end{align}
As a result, based on the regression fits presented in this work, we expect BHEX to be able to use the normalized transverse shift to constrain the spin of \m to within 
\begin{itemize}
    \item $9\%$ if the accretion-flow direction is prograde,
    \item {$25\%$} if the accretion-flow direction is retrograde,
    \item {$24\%$} if the accretion-flow direction is indistinguishable~\footnote{{Note: the spin error on the indistinguishable case is smaller than that of the retrograde case. While the errors on the mean parameters ($\sigma_{\bar{m}}$ and $\sigma_{\bar{b}}$) are larger in indistinguishable case, the best fit slope values are such that the relative errors obey ($\sigma_{\bar{m}}/\bar{m}$ and $\sigma_{\bar{b}}/\bar{m}$) are smaller for the indistinguishable case.}}.
\end{itemize}

The spin-constraint errors that we have estimated are based on the assumptions that the set of simulations we examined is representative of the possible accretion states of \m.
{As a conservative assumption, we took $R_{\rm high}$ to be distributed over a large range, 1-160; however, our simulation sample is limited by assuming perfect disk alignment and fixed observer inclination}.
Fortunately, there are observables that are correlated with accretion-state parameters{.} {EHTC analyses of jet power and horizon-scale polarization structure disfavor low $R_{\rm{high}}$ for the \m accretion flow {\citep{EHTC_M87_V,EHTC_M87_VIII}}. Additionally, accretion disk tilt is correlated with periodic jet motion ~\citep{Cui_M87Jet_LT_2023,Cui_M87Jet_LT_2025} and with horizon-scale image ring brightness distribution and variation \citep{Chatterjee_TiltedDisks_2020,White_TiltedDisks_2020}. Considering inclinations varying within the measured jet orientation uncertainty of $\pm 3^\circ$~\citep{Walker_jet}, the equatorial ring model suggests small changes in inclination should induce a minimal change in the spin-transverse shift relation}~\footnote{{Repeating the fit used to derive \autoref{eq:hERMFit} with the added assumption that the inclination is equally likely to lie anywhere in the range $\theta_o\in[14^\circ,20^\circ]$ only induces a small change in the inferred values of ${S_{\perp,\rm erm} (a_*)}$ and $\sigma_{\perp,\rm erm}$ for both $r_s/r_g$ range cases, but this extension requires investigation in the case of simulations.}}.
Repeating this analysis with a more representative simulation {sample is} expected to reduce uncertainties associated with astrophysical variation.

Updating the model used for shift extraction may offer another avenue to decrease the spin-constraint uncertainties we derive from relative astrometry.
The mF-ring model {herein} characterizes the geometric shape of the accretion-disk images {with an ellipse featuring two brightness modes}, which may be too constraining. Thus, models that admit more flexible shapes {and brightness modes} may be worth consideration. 
As an alternative to mF-ring-based characterization of accretion-disk-image shape, machine-learning-based algorithms have been developed to extract the shapes of the direct and secondary accretion-disk images \citep{Duong2026_in_prep}.

We note that in this work, for each simulation, we have fitted the brightness profile of a time-averaged image created using snapshots with a fixed camera origin. 
Hence, this analysis assumes the existence of a robust pipeline for {constructing and averaging} snapshots of the accretion flow {from interferometric data}. 
This pipeline is under independent development, but preliminary results suggest that straightforward image-alignment of \m snapshots taken over a full BHEX campaign provides an excellent approximation to the time-averaged images \citep{Zeng2026_in_prep}. {Repeating this analysis on the time-averaged images from synthetic observations will yield more refined estimates of the astrometry-based spin constraints possible using BHEX.}

Furthermore, our estimated spin-constraint uncertainties are based solely on relative astrometry of the photon ring. 
Other image features---including the secondary image's shape \citep{CardenasAvendano_n1_photon_ring_shape} and brightness profile \citep{Farah_ring_brightness,Bernshteyn_ring_asymmetry}, the inner shadow morphology \citep{Kearns_Inner_Shadow,Guan2026_in_prep}, the relative size \citep{Broderick_photon_ring_size,bhex_science} and polarization spiral pitch angles \citep{Palumbo_polarimetry_spin,Wong_spin_and_polarization} of the direct and secondary images---are known to correlate with spin. 
All these features will be measurable on the horizon-scale images and can be used simultaneously to produce stronger spin constraints. 

More promising horizon-scale spin-inference techniques arise by using Bayesian analysis to compare observed images to those produced using the geometric emission models \citep{jukebox}. 
In the context of these models, our work serves two purposes.
First, by focusing on the spin-driven normalized transverse shift and its derivation using simulations, we present a procedure for constraining black-hole spin that does not rely on geometric emission models.
Hence, this approach can therefore provide a complementary spin constraint when applied to the same images as the geometric-emission-model techniques.
Second, we provide a first-principles understanding of why spin physically drives the center shift and demonstrate that this effect persists in realistic systems.
In doing so, this work begins to demystify why geometric-emission-model inferences can accurately deduce black-hole spin, thereby validating their efficacy.

Lastly, the analyses presented here have demonstrated the power of the center shift in the specific context of \m assuming the large scale jet is aligned with the black hole's spin axis. To test this assumption and to enable spin measurements of Sgr A*, the other primary target for high-resolution horizon-scale imaging that will reveal the photon ring, it is valuable to develop techniques that do not require the observer inclination as an input.
Accordingly, this paper serves as the first in a series. 
In the next paper, we will develop methods for simultaneously constraining both spin and inclination by leveraging the displacement between the centers of the direct image and the photon ring.

\begin{acknowledgements}
\section*{acknowledgments}
The authors thank Michael Johnson and George Wong for discussions and comments on this work, and acknowledge the 2024 Photon Ring Workshop at Vanderbilt University where this work was initiated. 
{We also thank the anonymous referee for detailed feedback that improve our manuscript.}
This work was supported in part by the {National Science Foundation} (AST-2307887) and the Gordon and Betty Moore Foundation (Grant \#12987), and by the Black Hole Initiative through the Gordon and Betty Moore Foundation (Grant \#13526) and the John Templeton Foundation (Grant \#63445). 
The opinions expressed in this publication are those of the authors and do not necessarily reflect the views of
these Foundations.
\end{acknowledgements}

\appendix

\section{Center of equatorial emission rings on image plane}
\label{app:EqRingImages}

{Here we briefly review the appearance of image-plane projections of rings in the equatorial plane of the black hole spacetime on the image-plane of a distant observer. We present details for the locations of direct ($n=0$) and first lensed ($n=1$) projections of the equatorial rings.}  

The geometry around a black hole is described by the Kerr metric, which depends on the black-hole mass $M$ and angular momentum ${J}~{=}~{a_* M^2}$ in geometric {units} ${G}~{=}~{c}~{=}~{1}$. 
In Boyer-Lindquist coordinates $(t,r,\theta,\phi)$, the line element is
\begin{subequations}
\label{eq:Kerr}
\begin{gather}
	ds^2=-\frac{\Delta}{\Sigma}\pa{\ed t-a_* M\sin^2{\theta}\ed\phi}^2+\frac{\Sigma}{\Delta}\ed r^2+\Sigma\ed\theta^2
    +\frac{\sin^2{\theta}}{\Sigma}\br{\pa{r^2+a_*^2M^2}\ed\phi-a_*M\ed t}^2,\\
	\Delta=r^2-2Mr+a_*^2M^2,\quad
	\Sigma=r^2+a_*^2M^2\cos^2{\theta}.
\end{gather}
\end{subequations}
and the black hole's event horizon is located at
$r_{\rm H}= M\pa{1+\sqrt{1-a_*^2}}$.
The spin parameter $a_*$ ranges from ${a_*}~{=}~{0}$ (the non-spinning, {Schwarzschild} black hole) to ${a_*}~{=}~{1}$ (the maximally-spinning extremal black hole).

Photons traveling on the Kerr background have four-momentum
\begin{align}
	\label{eq:KerrMomentum}
	\frac{p}{E}={-\ed t +M\br{\frac{\pm_r\sqrt{\mathcal{R}(r)}}{\Delta(r)}\ed r\pm_\theta \sqrt{\Theta(\theta)}\ed\theta+\lambda\ed\phi}}\\
	\mathcal{R}=\br{\pa{\frac{r^2}{M^2}+a_*^2}-a_*\lambda}^2 -\Delta(r)\br{\eta+\pa{\lambda-a_*}^2},\quad
	\Theta=&\eta+a_*^2\cos^2 \theta -\lambda \cot^2 \theta
\end{align}
{where} $E$ is the energy of the photon, $\lambda$ is the angular momentum about the black hole's spin axis normalized by photon energy and black-hole mass, and $\eta$ is {the} Carter constant normalized by {the} photon energy and black-hole mass, all of which are conserved along a geodesic. 
Since the four-momentum is proportional to the photon's energy, the photon's trajectory is determined by its angular momentum and Carter constant.
{\autoref{eq:KerrMomentum} can be integrated to determine the trajectories of photons that make $n$ angular librations between a source point $\cu{r_s,\theta_s,\phi_s}$ and an observation point $\cu{r_o,\theta_o,\phi_o}$.} 

\begin{figure*}
    \centering
    \includegraphics[width=\linewidth]{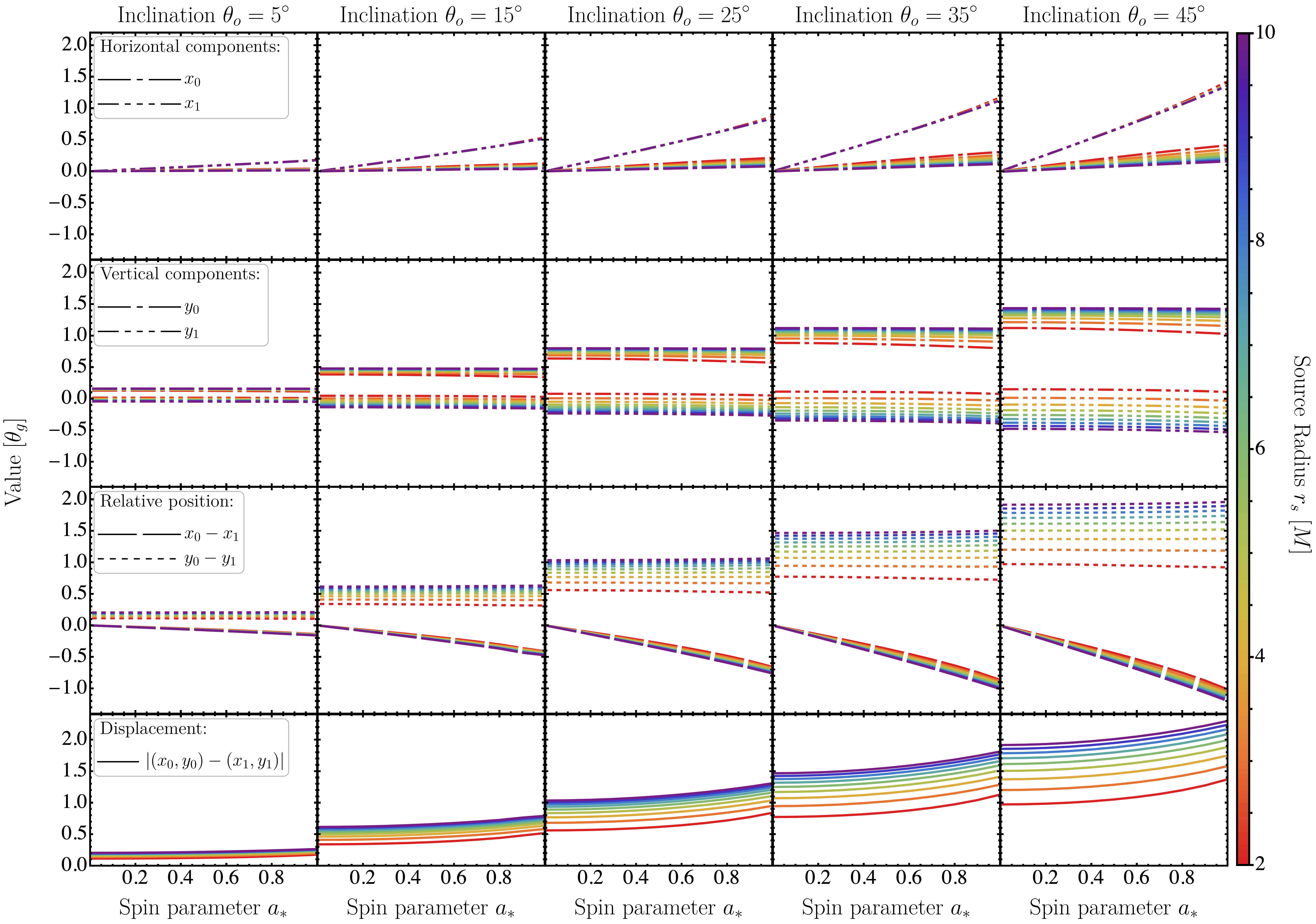}
    \caption{Behavior of the centers $(x_n,y_n)$ for the ${n}~{=}~{0}$ and ${n}~{=}~{1}$ images of equatorial source rings on the image plane \autoref{eq:ScreenCoords}. 
    We show {the horizontal components ($x_n$; top row), vertical components ($y_n$; second row), relative position ($({x_0}~{-}~{x_1},{y_0}~{-}~{y_1})$; third row), and displacement ($|{(x_0,y_0)}~{-}~{(x_1,y_1)}|$; bottom row) of the image centers
    as a function of spin (sub-plot horizontal axes), inclination (${\theta_o}~{\in}~{\cu{5^\circ,15^\circ,\dots,45^\circ}}$; columns) and source ring radii (${r_s}~{\in}~{\cu{2M,3M,\dots,10M}}$; colored curves)}. Quantities are shown in units of the mass-to-distance ratio ${\theta_g}~{=}~{M/D}$.
    }
    \label{fig:CentroidsApp}
\end{figure*}

{For equatorial source rings, we consider the locations (i.e., arc-length-weighted centroids denoted $(x_n,y_n)$) of direct (${n}~{=}~{0}$) and secondary (${n}~{=}~{1}$) images on the image plane of a distant (${r_o}~{=}~{D\gg M}$) observer \autoref{eq:ScreenCoords}.
We can heuristically describe the effects of spin and inclination on the image centers.}
When the black hole is non-spinning, symmetry requires that the {image centers} lie along the projection of the ring’s normal onto the image plane, with the {image center located off the image-plane origin} arising entirely from the observer’s inclination relative to the ring’s normal{, a purely general relativistic effect due to the asymmetric lensing of the near and far sides of the ring}
~\footnote{In the flat space case, a circle viewed of the normal appears as an ellipse with major axis ${A}~{=}~{r_s/r_o}$ and minor axis ${B}~{=}~{A\cos\theta_o}$, with center located a coordinate ${(x,y)}~{=}~{(0,(r_s\cos\theta_s)/(r_o \csc\theta_o}$)), which for equatorial rings simplifies to ${(x,y)}~{=}~{(0,0)}$.}.
If the black hole spins around the ring's normal direction, the image centers can shift off the projection of the normal (which now coincides with the black-hole spin axis). Photons arriving on the side of the spin axis that rotates toward the observer pass closer to the black hole than photons with the same angular momentum arriving from the side that rotates away from the observer. This effect arises from frame dragging, which makes it easier for photons to co-rotate with the black hole than to counter-rotate~\footnote{{This spin-induced $x$-coordinate shifting moves all $n$ images of equatorial rings rightward on the image plane, however the degree of shifting depends on the spin, inclination, and ring radius. This behavior persists in the large-$n$ limit, i.e., for the critical curve \citep{Bardeen_KerrGeodesics_1973}.}}. {In \autoref{fig:CentroidsApp}, we plot the image centers, as well as the relative location and displacement of the direct image relative to the secondary image, as a function of the system parameters $a_*$, $\theta_o$, and $r_s$.}

We find always ${x_1}~{>}~{x_0}~{>}~{0}$ for ${a_*}~{>}~{0}$. 
This is because the ${n}~{=}~{1}$ image comprises {photons} that travel on rays which spend more in the vicinity of the black hole and are thus more {affected} by the frame dragging.
Additionally, both $x_0$ and $x_1$ increase with spin and inclination and decrease with source radius. However, $x_1$ varies more quickly with spin and inclination than $x_0${, while} $x_0$ varies more quickly {with} source radius than $x_1$. In fact, $x_1$ is nearly independent of source radius {except when the spin and inclination are both large.} 
(See the {top} row of \autoref{fig:CentroidsApp}.) 
Together, these behaviors result in a center shift transverse to the projection of the black-hole spin axis, ${x_0}~{-}~{x_1}$, that is negative and decreasing in spin, inclination, and source radius. However, ${x_0}~{-}~{x_1}$ changes more rapidly in spin and inclination but changes slowly with source radius. 
(See the {dashed} lines in the {third} row of \autoref{fig:CentroidsApp}.) 

For ${a_*}~{>}~{0}$, ${y_0}~{>}~{y_1}$. 
Additionally, ${y_0}~{>}~{0}$ always; whereas, $y_1$ is typically negative but may be positive for a compact range of sufficiently small source radii. 
Furthermore, $y_0$ is increasing {with} inclination and source radius, while $y_1$ is decreasing in both. On the other hand, both $y_{0}$ and $y_{1}$ {decrease} with increasing spin. 
(See the {second} row of \autoref{fig:CentroidsApp}.) 
Together these behaviors produce a center shift parallel to the projection of the black-hole spin axis, ${y_0}~{-}~{y_1}$, that is positive, increases with inclination source radius, and decreases with spin. 
Notably, ${y_0}~{-}~{y_1}$ varies more rapidly with inclination and source radius than with spin. 
(See the {dotted} lines in the {third} row of \autoref{fig:CentroidsApp}.) 

{Finally, for} the equatorial source ring, we find that the magnitude of the center shift ${|(x_0,y_0)-(x_1,y_1)|} = {\sqrt{(x_0-x_1)^2+(y_0-y_1)^2}}$ grows monotonically with spin{,} inclination, {and} source radius $r_s${, highlighting that the magnitude of the center shift alone is inadequate for constraining spin.} (See the {bottom} row of \autoref{fig:CentroidsApp}.)

\section{{Weighted $R_{\rm high}$ fits}}
\label{app:Weights}

{Here we perform weighted least squares regressions on our sample of MAD simulation center shifts.}
We weight our {sample shift} measurements to compensate for non-uniform sampling of $R_{\rm high}$ in our simulation library, choosing weights to approximate an effective uniform prior on $R_{\rm {high}}$. 
For each of the ${N_R}~{=}~{6}$ sampled $R_{\rm high}$ values, ${R_j}~{\in}~{\cu{1,10,20,40,80,160}}$, the weights $w_j$ are assigned via one-dimensional Voronoi binning along the interval spanned by the {sampled} $R_{\rm high}$ values, $1$--$160$, and normalized such that the sum of all the weights is one. 

Specifically, each of the extremal sample value ($R_1$ and $R_{N_R}$) is associated with the interval extending from the sample value to the midpoint between it and its nearest neighbor. Each non-extremal sample value ($R_j$ with ${2}~{\leq}~{{j}}~{\leq}~{N_R-1}$) is associated with the interval extending halfway to its neighboring sample values. The weight for each sample value is taken to be the length of the associated interval normalized by the product of the total span (${R_{N_R}}~{-}~{R_1}$) and the number of unique spin values considered. 
Mathematically, this gives:
\begin{align}
    w_j&=\frac{\hat{w}_j}{2N_{{a_*}} \pa{R_{N_R}-R_{1}}},\quad \hat{ w_j}=\begin{cases}
        R_2-R_1, \quad &j=1\\
        R_{j+1}-R_{j-1}, & 2\leq j \leq N_R-1\\
        R_{N_R}-R_{N_{R}-1}, \quad &j=N_R
    \end{cases}.
\end{align}
For the collection of simulations considered herein, ${N_{a_*,+}}~{=}~{N_{a_*,-}}~{=}~{3}$ (for the prograde and retrograde cases) and ${N_{a_*,\pm}}~{=}~{5}$ (for the indistinguishable case).
{Hence, the} weights associated with each $R_{\rm high}$ are as follows:
\[\begin{tabular}{c|c c c c c c }
        $R_{\rm high}$ & \ 1 \ & \ 10 \ & \ 20 \ & \ 40 \ & \ 80 \ & \ 160\\
         \hline 
         \rule{0pt}{3.6ex} (weight) $\cdot N_{{a_*}}$ & $\dfrac{3}{106}$ & $\dfrac{19}{318}$ & $\dfrac{5}{53}$ &$\dfrac{10}{53}$ &$\dfrac{20}{53}$&$\dfrac{40}{159}$
\end{tabular}\]

{Performing weighted least squares regression, we obtain the parallel shift approximator (\autoref{eq:ParallelShiftApproximator}) 
\begin{align}
    \hat S_{\parallel,+}&=\br{(4.1\pm0.2)\pm 0.5}\cdot10^{-2},\\
    \hat S_{\parallel,-}&=\br{(4.6\pm0.2)\pm 0.5}\cdot10^{-2},\\
    \hat S_{\parallel,\pm}&=\br{(4.3\pm0.1)\pm 0.6}\cdot10^{-2},
\end{align}
for the prograde, retrograde, and indistinguishable cases (comparable to  \autoref{eq:vProFit}, \autoref{eq:vRetFit}, and \autoref{eq:vBothFit}, respectively); and, 
the transverse shift approximator (\autoref{eq:TransverseShiftApproximator}),
\begin{align}
    \hat S_{\perp,+}&=\br{(3.8\pm0.2)|a_*| +(0.3\pm0.1)\pm 0.2}\cdot10^{-2},
    \\
    \hat S_{\perp,-}&=\br{(2.4\pm0.3)|a_*| +(0.1\pm0.2)\pm0.4}\cdot10^{-2},
    \\
    \hat S_{\perp,\pm}&=\br{(3.1\pm0.4)|a_*| +(0.2\pm0.3)\pm0.7}\cdot10^{-2},
\end{align}
for the prograde, retrograde, and indistinguishable cases (comparable to  \autoref{eq:hProFit}, \autoref{eq:hRetFit}, and \autoref{eq:hBothFit}, respectively). Finally, considering these values, \autoref{eq:RelAstro}, and the relative astrometry bound of $\sigma_{\rm r.a.}\lesssim0.1\mu{\rm as}$, we estimate BHEX can use the transverse shift to constrain the spin parameter of \m to within 10\%, 26\%, and 29\% when the accretion-flow rotation sense is prograde, retrograde, or indistinguishable.
}

\section{center shifts for SANE simulations}
\label{app:SANE}
\begin{figure*}
    \centering
    {\Large SANE Simulations {at} $\theta_o=17^\circ$}\\
    \includegraphics[width=.497\linewidth]{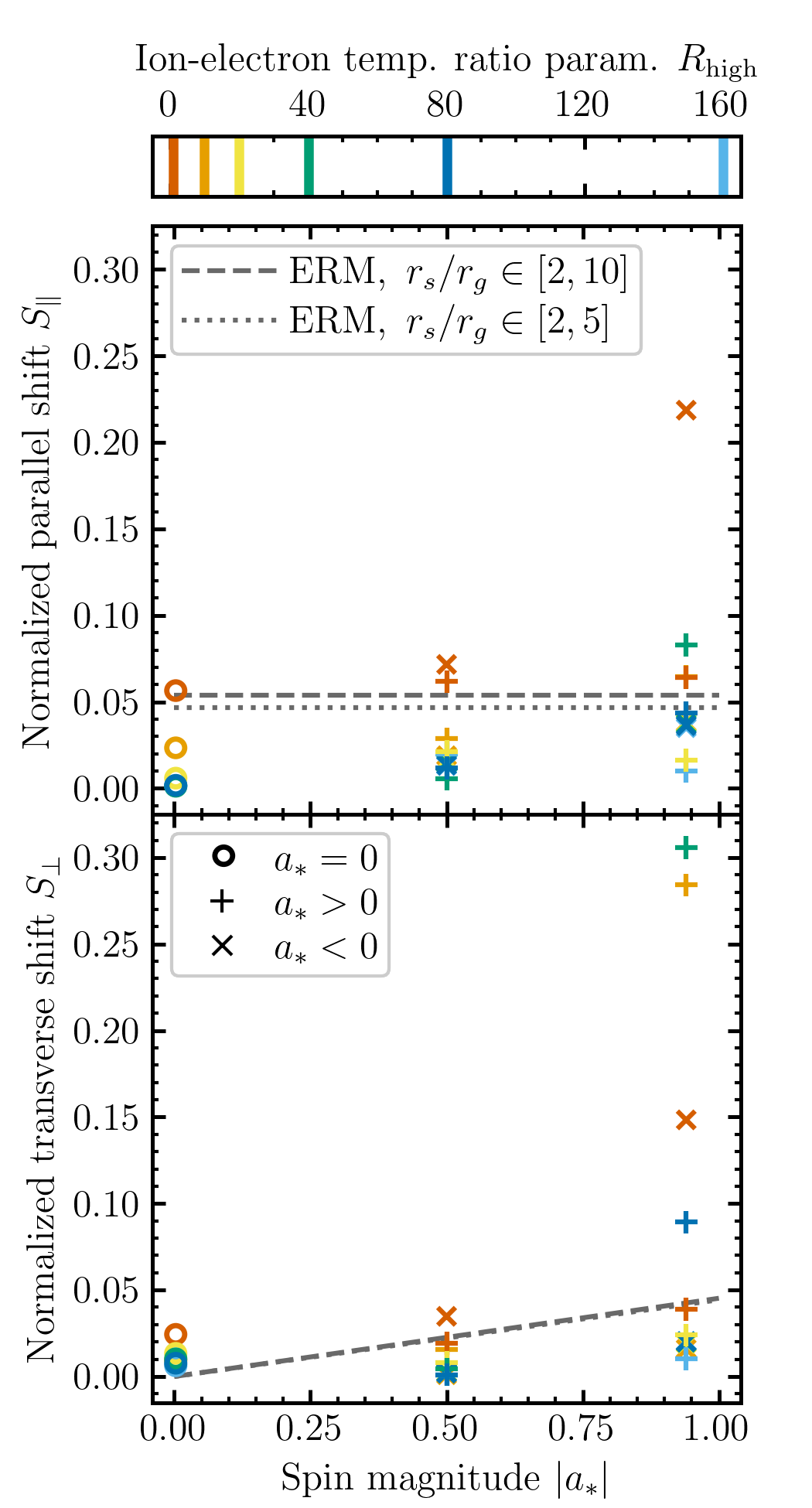}
    \includegraphics[width=.497\linewidth]{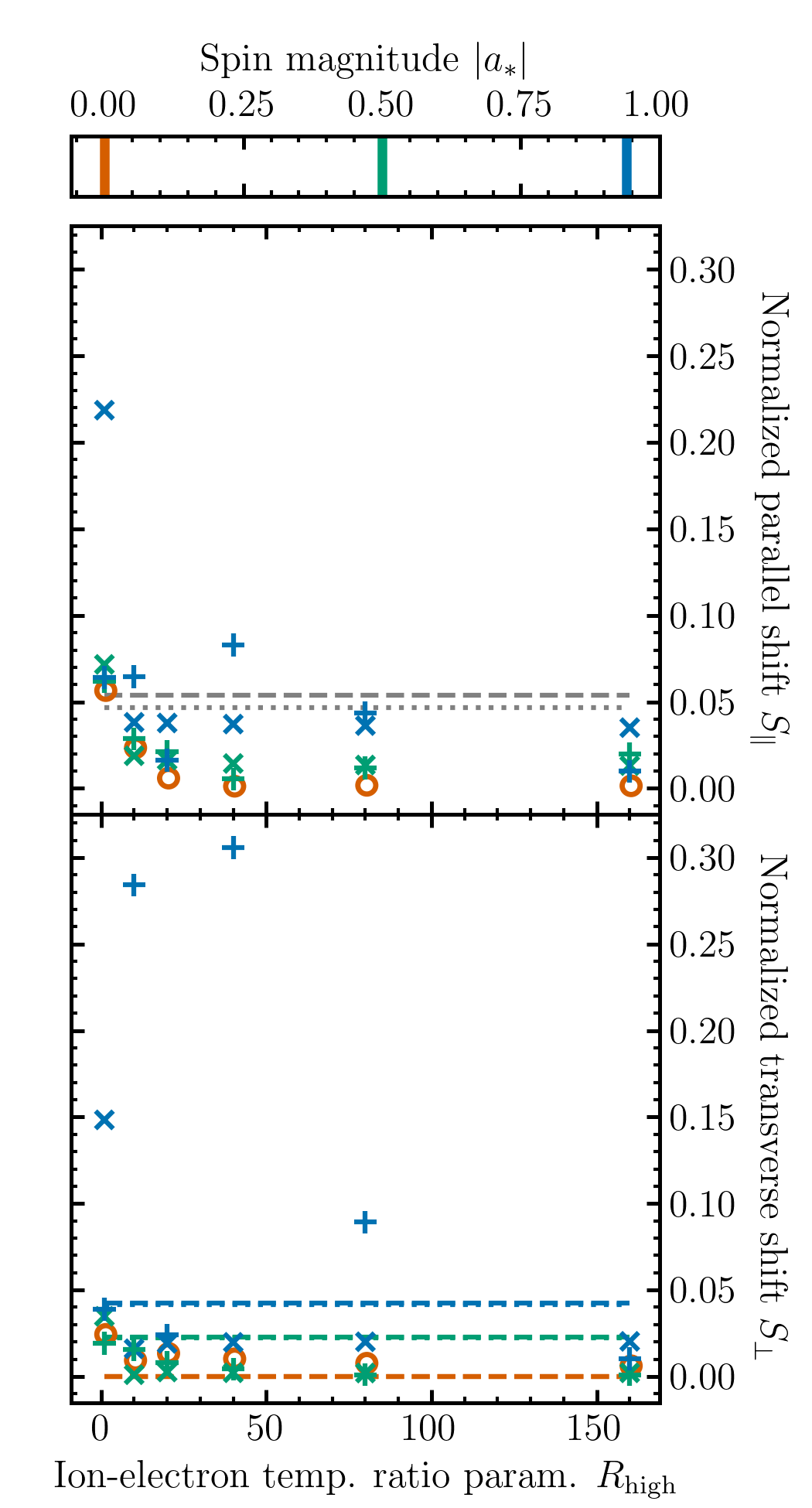}
    \caption{
    {Normalized shifts in images} of GRMHD SANE simulations viewed at ${\theta_o}~{=}~{17^\circ}$. The simulation shifts (scattered points) are compared to the {best-fit} mean shift values {of} the equatorial ring model {($\bar{S}_{\parallel,\rm{erm}}$ \autoref{eq:vERMFit} and $\bar{S}_{\perp,\rm{erm}}$ \autoref{eq:hERMFit}; lines)}. The coloring, marker-style, and line-style conventions match those used for the MAD simulations in \autoref{fig:MAD_GRMHD}.} 
    \label{fig:SANE_GRMHD}
\end{figure*}

For comparison to the set of {MAD simulations} examined in \autoref{sec:GRMHD}, we present the center shifts (\autoref{eq:CenterShifts}) extracted from simulations of \m-like {black-hole} systems with {SANE} accretion flows.
{The SANE sample consists of synthetic images from the \cite{Palumbo_Wong_2022} library (with parameters ${a_*}~{\in}~{\cu{0,\pm0.5,\pm0.94}}$, ${R_{\rm high}}~{\in}~{\cu{1,10,20,40,80,160}}$, and fixed ${R_{\rm low}}~{=}~{1}$)}, from which we extract the center shift using the mF-ring model procedure {outlined} in \autoref{sec:GRMHD}. 

The center shifts appear as scattered points in \autoref{fig:SANE_GRMHD}, where they are compared to the predicted shifts derived in \autoref{sec:ERM}. 
Unsurprisingly, the SANE-simulation center shifts can deviate drastically from the equatorial-ring-model predictions, in contrast to the MAD-simulation center shifts, which exhibit behavior similar to the equatorial-ring model. 
SANE simulations tend to be dominated by emission at higher scale heights, whereas MAD simulations tend to have equatorially dominated emission.

The equatorial ring model is {ill-suited} for modeling SANE-simulation center shifts {which feature significant amounts of off-equatorial emission~\citep{Emami_Emissivities_2023}}. 
As such, a different model is required to capture the general trends in SANE-simulation center shifts. 
Fortunately, MAD and SANE simulations often present morphologically distinct brightness profiles, making them discernible in high-resolution horizon-scale images. 
SANE-simulation images tend to exhibit an ${n}~{=}~{0}$ brightness profile that peaks inside the photon ring, near the inner shadow. By contrast, MAD-simulation images tend to exhibit an ${n}~{=}~{0}$ brightness profile more evenly distributed around the photon ring (for example, see Figure~2 in \cite{jukebox}). {Furthermore, other observables distinguish the MAD and SANE accretion states, including jet power and horizon-scale polarization structure, which have been used to determine that \m favors the MAD state~\citep{EHTC_M87_V,EHTC_M87_VIII}.}

\newpage

\bibliography{eht, references}

\end{document}